\documentclass[%
 reprint,
superscriptaddress,
 amsmath,amssymb,
 aps,
prb,
]{revtex4-2}

\setcitestyle{super,sort&compress}

\usepackage{microtype}
\usepackage{amsmath}
\usepackage{graphicx}
\usepackage{dcolumn}
\usepackage{bm}
\usepackage{adjustbox} 

\usepackage{color, hyperref}
\definecolor{mygrey}{gray}{0.80}
\definecolor{darkblue}{RGB}{8,81,156}
\hypersetup{colorlinks,breaklinks,
            linkcolor=darkblue,urlcolor=darkblue,
            anchorcolor=darkblue,citecolor=darkblue}

\definecolor{super-dark-green}{RGB}{0,69,41}
\definecolor{super-dark-purple}{RGB}{63,0,125}
\definecolor{super-dark-blue}{RGB}{8,48,107}
\definecolor{super-dark-red}{RGB}{165,0,38}

\begin{document}


\title{Sensitivity of Nucleation Thermodynamics and Kinetics to the Treatment of Long-Range Interactions}




\author{Fernanda Sulantay Vargas}
 \email{fernanda.vargas@yale.edu}
\affiliation{%
 Department of Chemical and Environmental Engineering, Yale University, New Haven, CT 06520, USA
}%
\author{Kimia Sinaeian}
 \email{kimia.sinaeian@yale.edu}
\affiliation{%
 Department of Chemical and Environmental Engineering, Yale University, New Haven, CT 06520, USA
}%
\author{Amir Haji-Akbari}
\email[Corresponding author: ]{amir.hajiakbaribalou@yale.edu}
\affiliation{%
 Department of Chemical and Environmental Engineering, Yale University, New Haven, CT 06520, USA
}%
\affiliation{%
 Quantitative Biology Institute, Yale University, New Haven, CT 06520, USA
}%
\affiliation{
Wu Tsai Institute, Yale University, New Haven, CT 06520, USA
}
\date{\today}

\begin{abstract}
\noindent
{\bf Abstract:}
Nucleation rates are exponentially sensitive to the thermodynamic driving force and can therefore depend strongly on the treatment of long-range intermolecular interactions. Here, using the Lennard--Jones (\textsc{Lj}) system as a benchmark, we combine molecular dynamics (\textsc{Md}) simulations, jumpy forward-flux sampling (\textsc{jFfs}), and free energy calculations to quantify the effect of potential truncation on melting thermodynamics, homogeneous crystal nucleation kinetics, and computational cost. Within the cutoff-radius range $2.5\sigma\le r_c\le 6\sigma$, the melting temperature at zero pressure varies by approximately $11\%$, while the nucleation rate changes by approximately ten orders of magnitude. By invoking classical nucleation theory (\textsc{Cnt}), we show that this pronounced kinetic sensitivity originates primarily from cutoff-induced changes in the chemical potential difference between the liquid and crystalline phases. Building on this observation, we develop a \textsc{Cnt}-based framework for extrapolating finite-cutoff rates to the full-potential limit and for estimating the expected rate deviations at other cutoff radii and temperatures. These findings also provide a systematic basis for cutoff selection: the optimal cutoff should minimize computational cost while keeping the deviation from the full-potential rate within acceptable bounds. At $kT/\epsilon=0.5$, $r_c=4\sigma$ provides a reasonable compromise according to these criteria. We further demonstrate that conventional homogeneous tail corrections do not offer a reliable alternative, as they cannot consistently account for the liquid, crystalline, and interfacial environments present during nucleation. Our findings highlight the need to specify and validate the truncation scheme as an integral component of force-field development in simulations of nucleation and other interfacial phase transitions within inhomogeneous environments.
\end{abstract}

\maketitle


\section{Introduction}
\label{section:intro}

\noindent
Crystal nucleation is the process by which crystalline domains emerge within metastable supercooled liquids or supersaturated solutions and often constitutes the rate-limiting step in crystallization. Characterizing the kinetics and mechanism of nucleation is therefore essential for understanding and controlling processes ranging from cloud microphysics and cryopreservation to food processing and advanced manufacturing.\cite{SossoChemRev2016} The central quantity in such analyses is the nucleation rate, which can depend exponentially on thermodynamic conditions, force-field parameters and implementation details. In recent decades, molecular simulations, often combined with advanced sampling techniques, have made it possible to access broader ranges of nucleation rates and obtain detailed mechanistic insights.\cite{tenWoldeJCP1996, AuerNature2001, FilionJCP2010, Li2011HomogeneousWater,  HajiAkbariPCCP2014, EspinosaJCP2015, HajiAkbariPNAS2015, HajiAkbariPNAS2017, JiangJCP2018p, YuanJPCB2023} However, quantitative rate prediction remains challenging because classical force fields often fail to reproduce key thermodynamic properties, such as chemical potential differences between phases and solid--liquid interfacial free energies. Even small deviations in these quantities can propagate into errors of many orders of magnitude in the predicted rate. It might therefore be necessary to apply \emph{post hoc} thermodynamic corrections to the computed nucleation rates.\cite{HajiAkbariPNAS2015}

While the sensitivity of crystal nucleation rates and pathways to the employed force field has received considerable attention,\cite{MendelevPhilosMag2008, EspinosaJCP2014, GianettiPCCP2016, JiangJCP2018, SanchezJCP2022, LauricellaJCP2023, MinhJPCB2025, SandbergJCP2025} their susceptibility to other implementation details, including system size,\cite{Hussain2021HowNucleation, Hussain2022HowNucleationb} thermodynamic ensemble,\cite{KarmakarJCTC2019, MonteroJPC2020} hydrodynamic effects,\cite{FiorucciJCP2020} and the employed computational method\cite{FilionJCP2010} and order parameter,\cite{BlowJCP2023, GispenJCP2024, Domingues2024DivergenceCrystals,Sinaeian2025} remains far less explored. One particularly underexamined aspect is the choice of cutoff distance for interatomic potentials that vanish  asymptotically. Truncating such potentials at a finite distance is commonly done to reduce computational cost, yet the choice of cutoff is often treated as a secondary consideration, as reasonable variations typically induce only modest changes in thermodynamic properties.\cite{CareJCP2007, IngebrigtsenPRX2012} Given the strong sensitivity of nucleation rates to thermodynamic driving forces, however, it is unclear whether such seemingly minor changes can lead to substantial variations in the predicted nucleation kinetics.

The Lennard--Jones (\textsc{Lj}) potential\cite{Jones1924OnTemperature}  is the most commonly used model for representing short-range repulsions and dispersion attractions in simple fluids, and has long served as a benchmark for testing molecular simulation techniques.\cite{RosenbluthJCP1954} In its standard 12--6 form, the repulsive term is steep and short-ranged, whereas the attractive term decays algebraically. As a result, the full potential is formally long-ranged in the sense that it does not remain identically zero beyond any finite separation. However, in practice, it is replaced with a finite-range version by truncating the potential at a cutoff radius $r_c$. To ensure numerical stability and physical fidelity, truncation is sometimes accompanied by strategies such as shifting,\cite{Weeks1971RoleLiquids} force-shifting,\cite{StoddardPRA1973} switching, or smoothing near the cutoff.\cite{BrooksJComputChem1983,SteinbachJComputChem1994} A related strategy is to supplement the configurational energy and pressure with analytical tail corrections when the potential is truncated without shifting.\cite{WoodJCP1957}  A particularly common choice is $r_c\approx2.5\,\sigma$,\cite{ToxvaerdJCP2011, LautenschlaegerFluidPhaseEquil2019} at which the magnitude of the \textsc{Lj} potential is only $\approx1/60$ of the well depth. This cutoff is computationally convenient and is often adequate for structural and bulk properties. However, its widespread use does not imply that quantities sensitive to small thermodynamic differences are similarly converged. In particular, it remains unclear whether $r_c=2.5\,\sigma$ provides an appropriate balance between accuracy and computational efficiency for crystal nucleation rate calculations.

Consistent with this concern, the physical consequences of different treatments mentioned above have been documented for a broad range of thermodynamic and kinetic properties of the \textsc{Lj} system, including virial coefficients,\cite{ShaulCCCC2010} phase coexistence,\cite{Smit1992PhaseFluids,Trokhymchuk1999ComputerAnswers,VrabecMolPhys2006, WangPCCP2024, MoroPCCP2024} interfacial properties,\cite{NijmeijerJCP1988,Trokhymchuk1999ComputerAnswers,ShenJPCB2007} transport coefficients,\cite{TakahashiJCP2007,BugelChemPhys2008,LautenschlaegerFluidPhaseEquil2019} and nucleation kinetics.\cite{Santra2008Gas-liquidSystem,WangPCCP2020} The melting temperature is particularly sensitive to such details because even small differential changes in the chemical potentials of the solid and liquid phases can produce considerable shifts in their coexistence temperature. Indeed, \textsc{Lj} melting curves reported in the literature differ by as much as 10\%.\cite{Mastny2007MeltingPotential} Systematic studies using a range of computational approaches have demonstrated a pronounced dependence of the melting temperature on truncation details, particularly the cutoff radius, thereby explaining at least part of the apparent variation in the literature.\cite{MorrisJCP2002,Mastny2007MeltingPotential,Ahmed2010EffectFluids,SchultzJCP2018}

Such variations can be particularly consequential in crystal nucleation studies, as the nucleation rate is especially sensitive to the thermodynamic driving force. Most importantly, classical nucleation theory (\textsc{Cnt}) predicts that the rate is exponentially dependent on the nucleation barrier, which itself is a function of the thermodynamic driving force and the two-phase interfacial tension.\cite{Volmer1926KeimbildungGebilden, Becker1935KinetischeDampfen, Turnbull1949RateSystems} However, the focus has been primarily on how truncation impacts the thermodynamics of freezing, with no systematic analysis of how the cutoff radius affects the kinetics of crystal nucleation. 

Addressing this gap also requires determining whether analytical tail corrections can recover the omitted attractive interactions in nucleation simulations.\cite{WoodJCP1957} Such corrections are well established for homogeneous bulk fluids, where the structure beyond $r_c$ can often be approximated by its bulk limit. However, their application is less straightforward in inhomogeneous systems, because the local density varies spatially. More elaborate treatments have therefore been developed for interfaces,\cite{ChapelaFaradayTrans1977,LotfiMolSimul1990,BlokhuisMolPhys1995,JanecekJPCB2006} adsorption,\cite{RowleyJComputPhys1978} and porous materials.\cite{Jablonka2019} Since the heterogeneity that emerges during nucleation is both irregular and time-dependent, it remains unclear whether a homogeneous tail correction improves the representation of long-range attractions consistently throughout the system or instead introduces spatially dependent errors.

Here, we combine molecular dynamics (\textsc{Md}) simulations, free energy calculations, and jumpy forward-flux sampling (\textsc{jFfs})\cite{Haji-Akbari2018Forward-fluxParameters} to quantify how the \textsc{Lj} cutoff radius affects crystal nucleation. We examine cutoff radii from $2.5\,\sigma$ to $6\,\sigma$ and determine their impact on melting thermodynamics, nucleation kinetics, and computational cost. By invoking \textsc{Cnt}, we show that the strong cutoff dependence of the nucleation rate arises primarily from changes in the thermodynamic driving force, and use this relationship to extrapolate finite-cutoff rates to the full-potential limit. We further develop a framework for estimating cutoff-induced rate deviations at different thermodynamic state points, providing a systematic basis for selecting a cutoff that balances computational efficiency against accuracy. By these criteria, $r_c\approx4\,\sigma$ emerges as a practical choice for nucleation calculations at $kT/\epsilon=0.5$. Finally, we show that conventional homogeneous tail corrections cannot consistently recover the omitted long-range interactions across the structurally distinct liquid, interfacial, and crystalline environments that emerge during nucleation.

\begin{figure*}
    \centering
    \vspace{-20pt}
    \includegraphics[width=.6039\textwidth]{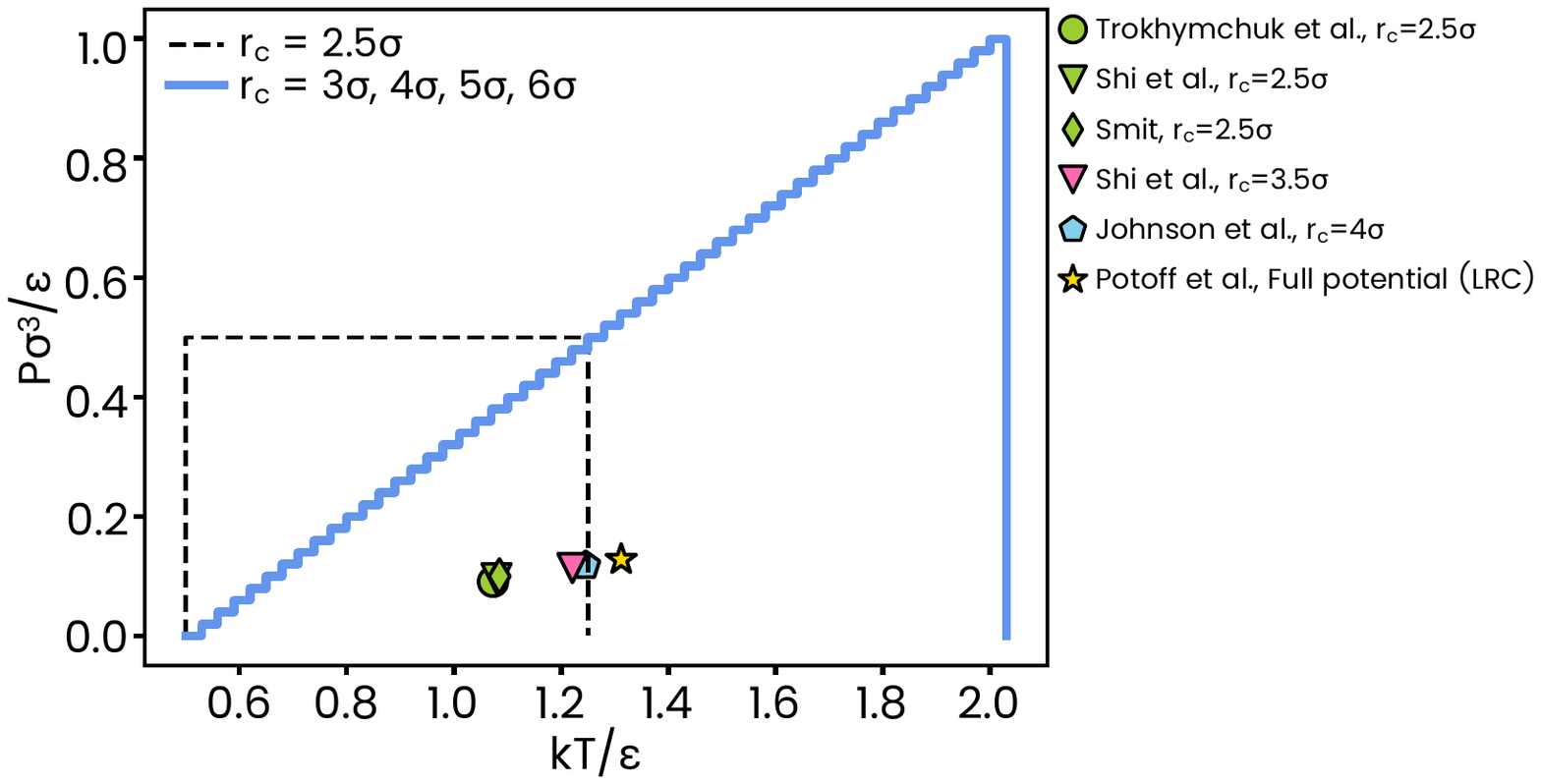}
    \vspace{-40pt}
    \caption{\label{fig:path} Reversible paths in the $p$--$T$ plane used for thermodynamic integration of the liquid \textsc{Lj} system. The dashed path is employed for $r_c=2.5\sigma$, whereas the solid path is used for $r_c=3\sigma$, $4\sigma$, $5\sigma$, and $6\sigma$. Each path circumvents the liquid--vapor critical point associated with the corresponding cutoff radius. Critical point locations were taken from Trokhymchuk \textit{et al.},\cite{Trokhymchuk1999ComputerAnswers}  Shi \textit{et al.},\cite{Shi2001HistogramFluids} and Smit\cite{Smit1992PhaseFluids} for $r_c=2.5\sigma$; Shi \textit{et al.}\cite{Shi2001HistogramFluids} for $r_c=3.5\sigma$; Johnson \textit{et al.}\cite{Johnson1993TheRevisited} for $r_c=4.0\sigma$; and Potoff and Panagiotopoulos\cite{Potoff1998CriticalMixture} for the full potential.}
    \vspace{-10pt}
\end{figure*}

The remainder of this manuscript is structured as follows. Section~\ref{section:methods} describes the simulation protocols and computational procedures utilized in this work. The results are presented in Section~\ref{section:results}, followed by a discussion in Section~\ref{section:force-field} of their implications for the development of classical force fields. Finally, Section~\ref{section:conclusions} summarizes the main conclusions and some broader observations.

\section{Methods}
\label{section:methods}

\subsection{Molecular Dynamics Simulations}\label{md_simulations}
\noindent
All \textsc{Md} simulations are conducted using \textsc{Lammps},\cite{Thompson2022LAMMPSScales} with all quantities reported in reduced \textsc{Lj} units. Unless otherwise specified, each system contains 6,912 particles in a fully periodic cubic simulation box, with average dimensions $l_x= l_y= l_z\approx19.87\,\sigma$.  Unless otherwise specified, simulations are performed in the isothermal isobaric (\textsc{Npt}) ensemble at $kT/\epsilon=0.5$ and $p\sigma^3/\epsilon=0$. Particles interact through a truncated and shifted \textsc{Lj} potential,\cite{Jones1924OnTemperature}
\begin{subequations}
\begin{eqnarray}
u_{\textsc{Lj}}^{\text{ts}} (r;r_c) &=& \left\{
\begin{array}{ll}
u_{\textsc{Lj}} (r) - u_{\textsc{Lj}} (r_c) & r\le r_c\\
0 & r>r_c
\end{array}
\right.,\\
u_{\textsc{Lj}} (r) & = & 4\epsilon\left[
\left(\frac{\sigma}{r}\right)^{12} - \left(\frac{\sigma}{r}\right)^{6}
\right],
\end{eqnarray}
\end{subequations}
with $\epsilon=1$, $\sigma=1$ and cutoff radii $r_{c}/\sigma=2.5$, 3, 4, 5, and 6. Following our earlier simulation protocol,\cite{Hussain2022HowNucleationb} the equations of motion are integrated using the velocity Verlet algorithm\cite{SwopeJCP1982} with a reduced time step of $\delta t^*=2.5\times10^{-3}$, where $t^*=(\epsilon/m\sigma^2)^{1/2}t$ and $m=1$ is the mass of each particle. Temperature and pressure are maintained using the Nos\'e--Hoover thermostat\cite{Nose1984AEnsemble,Hoover1985CanonicalDistributions} and Parrinello--Rahman barostat,\cite{Parrinello1981PolymorphicMethod} with coupling time constants of $10^2\delta t$ and $10^3\delta t$, respectively. For $r_c=2.5\sigma$, we use the basin-equilibrated configurations generated in our earlier study.\cite{Haji-Akbari2018Forward-fluxParameters} Simulations at the larger cutoff radii are initialized from these configurations, which are first equilibrated for an additional $4\times10^5$ \textsc{Md} steps using the new cutoff before being used in the subsequent rate and free energy calculations. All computational benchmarks are conducted on a dual-socket node equipped with two 18-core Intel Xeon Gold 6240 processors (Cascade Lake; 192~GB of memory).

\subsection{Free Energy Calculations}
\label{section:free-energy-calcs}
\noindent
We compute the free energy difference between the supercooled liquid and the face centered cubic (\textsc{Fcc}) crystal using thermodynamic integration combined with a variant of the Frenkel–Ladd method,\cite{Frenkel1984NewSpheres,Haji-Akbari2011PhaseTetrahedra} following the protocol detailed in our earlier study.\cite{SulantayVargas2026} Only the main steps and the modifications introduced for the different cutoff radii are summarized here.

To obtain the chemical potential of the supercooled liquid, we reversibly transform it into an ideal gas along a sequence of isotherms and isobars that bypasses the liquid--vapor critical point, as illustrated in Fig.~\ref{fig:path}. Changes in chemical potential along these segments are evaluated from
\begin{eqnarray}
\mu(p_2,T)-\mu(p_1,T)
&=&
\int_{p_1}^{p_2}\langle v\rangle\,dp,
\label{eq:P-int}\\
\frac{\mu(p,T_2)}{T_2}-\frac{\mu(p,T_1)}{T_1}
&=&
-\int_{T_1}^{T_2}\frac{\langle h\rangle}{T^2}\,dT,
\label{eq:T-int}
\end{eqnarray}
where $\langle v\rangle$ and $\langle h\rangle$ are the average specific volume and configurational enthalpy per particle, respectively, with the latter defined as $\langle h\rangle = \langle u\rangle+p\langle v\rangle$. Therefore, $\mu$ denotes the configurational contribution to the chemical potential, so that the kinetic energy contribution is excluded consistently. The corresponding integrals are evaluated using the interpolation scheme described in Ref.~\citenum{SulantayVargas2026}. All liquid free energy calculations are conducted using systems of 6,912 particles.

Since the location of the liquid--vapor critical point depends on the cutoff radius and truncation protocol, separate integration paths are constructed for $r_c/\sigma\in\{2.5,3,4,5,6\}$. These paths are chosen using literature estimates of the critical points of truncated and shifted \textsc{Lj} systems with comparable cutoff radii, as summarized in Fig.~\ref{fig:path}.\cite{Smit1992PhaseFluids, Johnson1993TheRevisited, Potoff1998CriticalMixture, Trokhymchuk1999ComputerAnswers, Shi2001HistogramFluids}

To compute the free energy of the \textsc{Fcc} crystal, we reversibly couple it to an Einstein crystal by introducing a harmonic restraining potential,
\begin{eqnarray}
\mathcal{H}(\mathbf{r}^n;\gamma)
=
\mathcal{U}_{\textsc{LJ}}(\mathbf{r}^n)
+
\frac{\gamma}{2\sigma^2}
\sum_{i=1}^{n}
|\mathbf{r}_i-\mathbf{r}_{i,0}|^2,
\end{eqnarray}
where $\mathbf{r}_{i,0}$ denotes the ideal lattice position of particle $i$. The Helmholtz free energy per particle of the unrestrained crystal is then obtained from
\begin{eqnarray}
f_x(T;0)
=
f_x(T;\gamma_{\max})
-
\frac{1}{2\sigma^2}
\int_0^{\gamma_{\max}}
\left\langle
|\mathbf{r}-\mathbf{r}_0|^2
\right\rangle_\gamma\,d\gamma,\notag\\
&&
\label{eq:FL-integral}
\end{eqnarray}
where $f_x(T;\gamma_{\max})$ is evaluated using the analytical Einstein crystal limit and $\left\langle
|\mathbf{r}-\mathbf{r}_0|^2\right\rangle_\gamma$ is the per-particle mean-squared displacement at fixed $\gamma$. The corresponding chemical potential is given by
$\mu_x=f_x+p\langle v_x\rangle_p$.

\begin{figure}
\centering
\includegraphics[width=0.2983\textwidth]{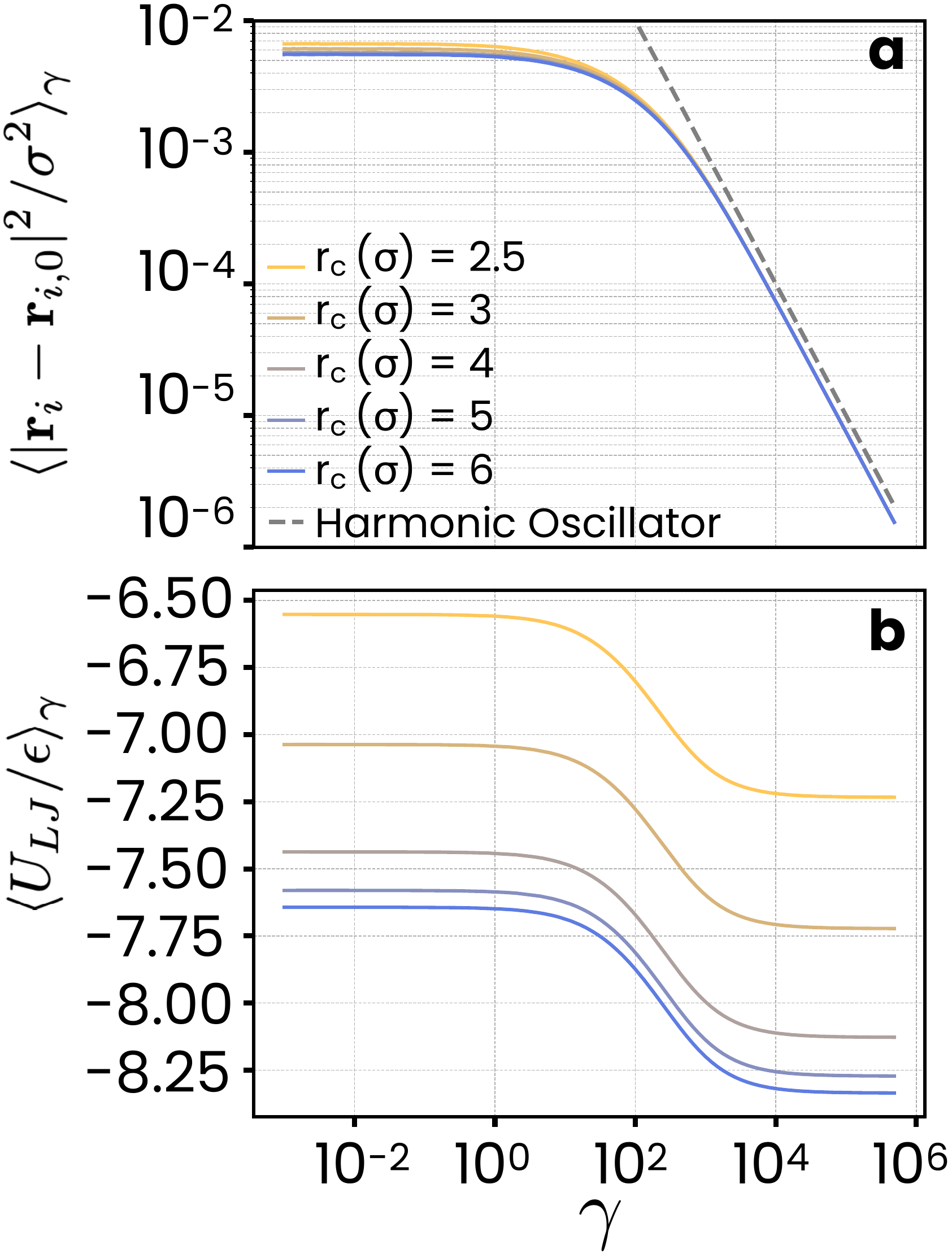}
\vspace{-10pt}
\caption{(a) Mean-squared displacement and (b) average potential energy as a function of the coupling parameter $\gamma$, for different cutoff radii. The dashed line in (a) corresponds to the slope of the analytical large-$\gamma$ limit for non-interacting harmonic oscillators.
\label{fig:MSD-AvgU}}
\vspace{-10pt}
\end{figure}

For each cutoff radius, we first conduct \textsc{Npt} simulations of an \textsc{Fcc} crystal containing 13,500 particles to determine its zero-pressure density at $kT/\epsilon=0.5$. The resulting reduced densities are given in Table~\ref{table:LJ_homrates}.  Canonical ensemble simulations are subsequently performed at these densities for 201 logarithmically spaced values of $\gamma$ between $10^{-3}$ and $4.85\times10^5$. The upper limit is chosen such that the average \textsc{Lj} potential energy approaches its ideal lattice value (Fig.~\ref{fig:MSD-AvgU}b) and the mean-squared displacement follows the expected harmonic behavior (Fig.~\ref{fig:MSD-AvgU}a). An additional \textsc{Nvt} simulation at $\gamma$=0 is performed to evaluate the mean-squared displacement per particle at the lower integration bound. To avoid numerical instabilities associated with the Nos\'e--Hoover thermostat at large values of $\gamma$,\cite{Martyna1992Nose-HooverDynamics} we instead use a Langevin thermostat with a reduced time step of $\delta t^*=10^{-4}$.

After computing the chemical potentials of the liquid and crystal at $kT/\epsilon=0.5$ and $p\sigma^3/\epsilon=0$, we conduct additional \textsc{Npt} simulations to extend both quantities over $kT/\epsilon\in[0.5,0.73]$ at zero pressure. The resulting chemical potential differences are used to determine the thermodynamic driving force for crystallization at each cutoff radius.

\subsection{Rate Calculations}
\noindent
We compute nucleation rates using \textsc{jFfs},\cite{Haji-Akbari2018Forward-fluxParameters} a generalization of conventional forward-flux sampling\cite{Allen2006SimulatingSystems, Hussain2020StudyingOutlook} applicable to order parameters that do not exhibit continuous temporal evolution. We use the size of the largest crystalline nucleus as the order parameter, with crystalline particles identified using neighbor-averaged\cite{Lechner2008AccurateParameters} Steinhardt bond order parameters.\cite{Steinhardt1983Bond-orientationalGlasses}

For each cutoff radius, the supercooled liquid basin is sampled using 96 independent \textsc{Md} trajectories with a combined duration ranging from $8.5\times10^5$ to $1.2\times10^6\,\sigma\sqrt{m/\epsilon}$. At least 4,000 successful crossings are collected at every subsequent \textsc{jFfs} milestone. The resulting nucleation rates are reported in Table~\ref{table:LJ_homrates} with the error bars estimated using the approach outlined in Ref.~\citenum{AllenJCP2006}. As established in our earlier finite-size analysis,\cite{Hussain2022HowNucleationb} a system containing 6,912 particles avoids significant finite-size artifacts for $r_c=2.5\sigma$. Since the critical nuclei are smaller at the larger cutoff radii considered here, this system size constitutes an even more conservative choice for those systems.

\subsection{Radial Distribution Functions}
\label{subsection:RDFcalc}
\noindent
Two sets of radial distribution functions (\textsc{Rdf}s), $g(r)$, are computed: one for the bulk supercooled liquid and another for the crossing ensemble at each \textsc{jFfs} milestone. In both cases, pair separations are evaluated up to $6\sigma$ by extending the neighbor and communication ranges used for the structural analysis.  For the bulk supercooled liquid, $g(r)$ is accumulated from independent conventional \textsc{Md} trajectories conducted separately at each interaction cutoff under the same thermodynamic conditions outlined in Section~\ref{md_simulations}. Each trajectory is equilibrated for $75\tau$, followed by a $1000\tau$ production period, with $\tau=\sigma\sqrt{m/\epsilon}$. Pair separations are binned into 200 spherical shells. For each \textsc{jFfs} milestone, $g(r)$ is evaluated by post-processing the saved crossing configurations without further propagation. Since fewer configurations are available for these ensembles, pair separations are binned into 160 spherical shells to assure adequate statistics within each shell. The resulting profiles are subsequently used to estimate the energetic contribution of the interactions omitted beyond the cutoff.

\section{Results and Discussion}
\label{section:results}

\noindent
\subsection{Sensitivity of Melting Temperature to \textsc{LJ} Cutoff Radius}
\label{section:meltingT_sensitivity}
\noindent

\begin{figure}
    \centering
    \includegraphics[width=0.3726\textwidth]{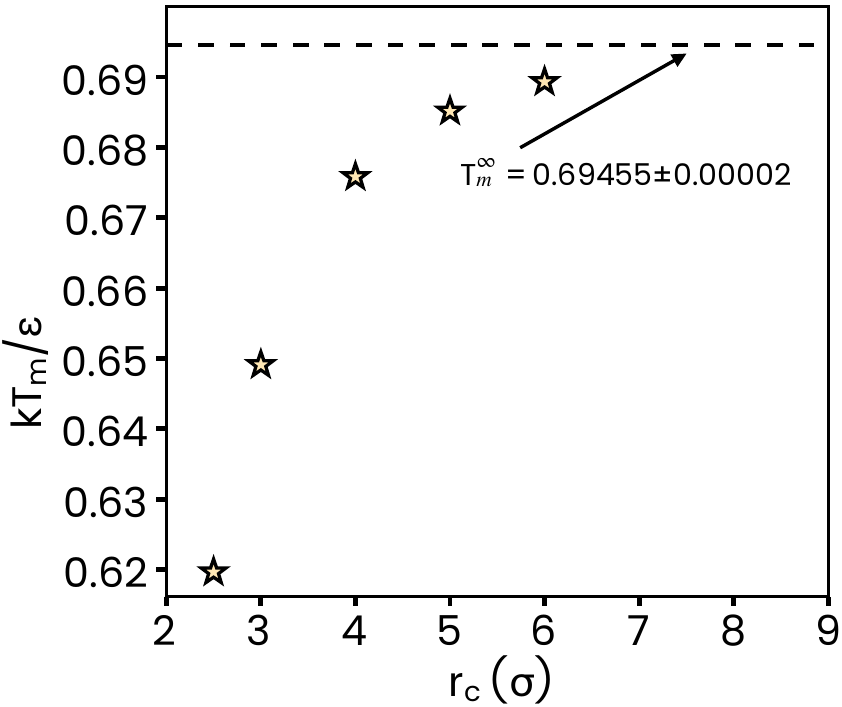}
\caption{\label{fig:TmEffects} Cutoff dependence of the melting temperature of the \textsc{Lj} system at $p\sigma^3/\epsilon=0$. For $r_c\ge4\,\sigma$, the melting temperatures lie near the full-potential value of $kT_m/\epsilon\approx0.6945$.\cite{Agrawal1995ThermodynamicCoexistence, SchultzJCP2018}}
    \vspace{-10pt}
\end{figure}

\noindent
To quantify the sensitivity of the melting temperature to the cutoff radius, we use the procedure described in Section~\ref{section:free-energy-calcs} to determine $T_m$ as a function of $r_c$ at zero pressure. The resulting melting temperatures are depicted in Fig.~\ref{fig:TmEffects}. The melting temperature exhibits a pronounced cutoff dependence at smaller $r_c$'s, including the commonly employed value of $r_c=2.5\sigma$. Increasing the cutoff radius from $2.5\sigma$ to $3\sigma$ and $4\sigma$ raises $T_m$ by $4.75\%$ and $9.07\%$, respectively. This substantial variation demonstrates that the omitted long-range attractions affect the relative thermodynamic stability of the liquid and the crystal and therefore cannot be regarded as a negligible contribution to melting. For $r_c\geq4\sigma$, the melting temperature changes only weakly with increasing $r_c$ and approaches the full-potential estimates reported by Agrawal \textit{et al.}\cite{Agrawal1995ThermodynamicCoexistence} and Schultz \textit{et al.}\cite{SchultzJCP2018} Thus, a cutoff radius of approximately $4\sigma$ is sufficient to almost recover the melting temperature of the full \textsc{Lj} potential, whereas the standard $2.5\sigma$ cutoff leads to a substantial underestimation. 

\begin{table}
    \centering
    \caption{Solid densities and homogeneous nucleation rates in the \textsc{Lj} system at $p=0$ and $kT/\epsilon=0.5$. Reported uncertainties correspond to 95\% confidence intervals.}
    \label{table:LJ_homrates}
    \begin{tabular}{c|c|c} 
        \hline\hline
        $r_{c} \left[\sigma\right]$ & $\rho_s\,[\sigma^{-3}]$ & $\log_{10}R_{\text{hom}} [\sigma^{-4}\epsilon^{1/2}m^{-1/2}]$\\ [0.5ex] 
        \hline
        2.5 & $0.9776$ & ~$-18.4488 \pm 0.0703$ \\ 
        3.0 & $0.9908$ &~$-12.8953 \pm 0.0508$ \\
        4.0 & $1.0005$ &~$-9.5074 \pm 0.0348$\\
        5.0 & $1.0041$ &$-8.6283 \pm 0.0347$\\ 
        6.0 & $1.0060$ & $-8.3078 \pm 0.0384$\\ [1ex] 
        \hline\hline
    \end{tabular}
\end{table}

The monotonic increase in $T_m$ observed here is consistent with the findings of Ahmed and Sadus,\cite{Ahmed2010EffectFluids} but contrasts with the oscillatory cutoff dependence reported by Mastny and de Pablo.\cite{Mastny2007MeltingPotential} However, we note that the latter calculations were conducted at a different pressure and employed an unshifted truncated potential together with analytical long-range corrections. Accordingly, the difference between our results and those of Mastny and de Pablo\cite{Mastny2007MeltingPotential}  should be interpreted in light of the distinct interaction potentials and thermodynamic conditions considered in the two studies.

\subsection{Sensitivity of Nucleation Kinetics to LJ Cutoff Radius}

\begin{figure*}
    \centering
    \vspace{-10pt}
    \includegraphics[width=0.5937\textwidth]{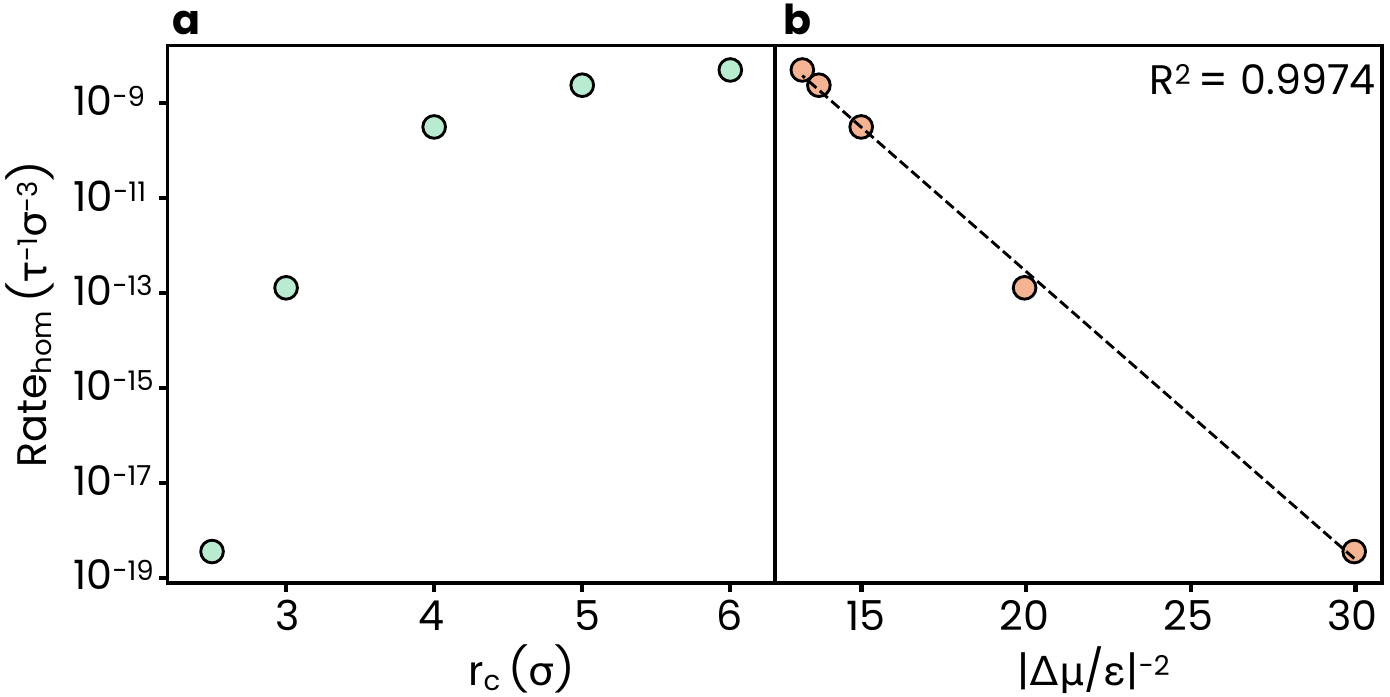}
    \vspace{-10pt}
    \caption{\label{fig:kinetic-eff} (a) Homogeneous nucleation rates in the \textsc{Lj} system at $p\sigma^3/\epsilon=0$ and $kT/\epsilon=0.5$ for different cutoff radii. The symbols correspond to actual rates computed from \textsc{jFfs} with error bars smaller than the symbols. (b) Relationship between  $\ln R_{\text{hom}}$ and $|\Delta\mu|^{-2}$, consistent with \textsc{Cnt} predictions.}
\end{figure*}

\noindent
After characterizing the effect of truncation on the thermodynamics of melting, we compute homogeneous nucleation rates at the same cutoffs. Figure~\ref{fig:kinetic-eff}a depicts the dependence of the  rate on the cutoff radius. The rate increases monotonically by approximately ten orders of magnitude as $r_c$ increases from $2.5\sigma$ to $6\sigma$. Most of this variation occurs at the shorter cutoff radii: increasing $r_c$ from $2.5\sigma$ to $4\sigma$ raises the rate by nearly nine orders of magnitude, whereas a further increase to $6\sigma$ changes it by only about one order of magnitude. Thus, the commonly employed cutoff of $2.5\sigma$ substantially suppresses nucleation, while $r_c=4\sigma$ marks the onset of a considerably weaker cutoff dependence. The analogous weakening in the sensitivity of the melting temperature to $r_c$ suggests that $4\sigma$ may serve as a practical minimum cutoff for nucleation rate calculations.

To understand the origin of this strong kinetic sensitivity, we invoke the predictions of classical nucleation theory.\cite{Volmer1926KeimbildungGebilden,Becker1935KinetischeDampfen,Turnbull1949RateSystems} Within \textsc{Cnt}, the homogeneous nucleation rate is given by
\begin{eqnarray}
R_{\text{hom}}
&=&
A_{\text{hom}}
\exp\left[
-\frac{\Delta G_{\text{hom}}^*}{kT}
\right],
\label{eq:rate-CNT}
\end{eqnarray}
where $A_{\text{hom}}$ is the kinetic prefactor and 
\begin{eqnarray}
\Delta G_{\text{hom}}^*
&=&
\frac{16\pi\gamma_{ls}^3}
{3\rho_s^2|\Delta\mu|^2}
\label{eq:deltaGstar}
\end{eqnarray}
is the nucleation barrier. Here, $\gamma_{ls}$ is the liquid--solid surface tension, $\rho_s$ is the number density of the crystal, and $|\Delta\mu|$ is the chemical potential difference between the solid and liquid,~i.e.,~the thermodynamic driving force for crystallization. Since all rate calculations are performed at the same temperature, the cutoff dependence of the nucleation barrier can arise through changes in $|\Delta\mu|$, $\gamma_{ls}$, and $\rho_s$. As mentioned in Section~\ref{section:free-energy-calcs}, $\rho_s$ changes by approximately 3\% upon altering $r_c$, potentially accounting for a $\approx6\%$ change in nucleation barrier, which is insufficient to explain the observed variations in rate. If we assume that $A_{\text{hom}}$ and  $\gamma_{ls}$ are both  weak functions of the cutoff radius, \textsc{Cnt} predicts a linear relationship between $\ln R_{\text{hom}}$ and $|\Delta\mu|^{-2}$:
\begin{eqnarray}
\ln R_{\text{hom}} &=& -\frac{B}{|\Delta\mu|^2} + C
\label{eq:linear-logR-dmusq}
\end{eqnarray}
 As depicted in Fig.~\ref{fig:kinetic-eff}b, the computed rates and chemical potentials exhibit a remarkably strong linear relationship consistent with Eq.~\eqref{eq:linear-logR-dmusq}, with $R^2=0.9974$. This observation indicates that the cutoff dependences of $A_{\text{hom}}$ and $\gamma_{ls}$ are comparatively weak over the range considered here. 
 The approximately ten-order-of-magnitude change in the nucleation rate can thus be attributed primarily to the dependence of the thermodynamic driving force, $|\Delta\mu|$, on the truncation cutoff. Although small variations in $A_{\text{hom}}$ and $\gamma_{ls}$ cannot be ruled out, they are not necessary to explain the dominant trend.

This finding parallels our earlier investigation\cite{GianettiPCCP2016} of a family of modified Stillinger--Weber\cite{StillingerPRB1985} liquids that include the monatomic water\cite{MolineroJPCB2009} (mW) model, in which increasing the tetrahedrality parameter changes the homogeneous nucleation rate by approximately 48 orders of magnitude, with most of this variation likewise attributable to changes in $|\Delta\mu|$. In that case, the thermodynamic driving force is altered by deliberately changing the tetrahedrality of the underlying force field. Here, a nominally technical choice-- the cutoff radius-- likewise alters the thermodynamic driving force, with a pronounced effect on the nucleation rate. These findings underscore how even modest changes in phase stability due to altering force-field parameters can translate into enormous changes in nucleation kinetics.

\begin{figure}
\centering 
 \includegraphics[width=0.4851\textwidth]{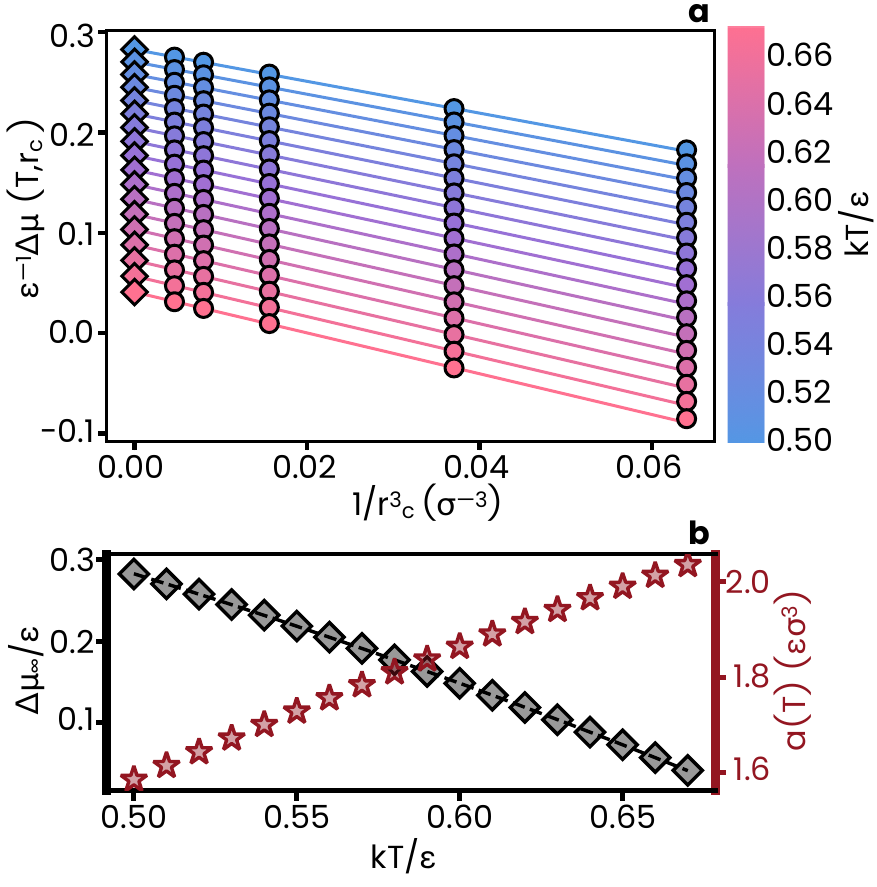}
 \caption{\label{fig:Delta-mu-rc}
 (a) Scaling of finite-cutoff $\Delta\mu$ with $r_c^{-3}$ at different temperatures. (b) The dependence of $\Delta\mu_{\infty}$ (gray) and $a(T)$ (red) on temperature. 
 }
\end{figure}

The linear relationship established in Eq.~\eqref{eq:linear-logR-dmusq} provides a route for estimating the nucleation rate in the full-potential limit. This is achieved by first extrapolating the thermodynamic driving force, $\Delta\mu=\mu_l-\mu_s$, at $r_c\rightarrow\infty$. For a pair potential whose slowest-decaying contribution is of the form $u(r)\sim r^{-m}$ ($m>3$), standard free energy perturbation and long-range tail correction arguments imply that the contribution from interactions omitted beyond $r_c$ will scale as $\int_{r_c}^{\infty}r^2u(r)\,dr\sim r_c^{3-m}$.\cite{ZwanzigJCP1954} For a truncated and shifted potential, an additional contribution arises from shifting the retained pair interactions by $-u(r_c)$. Since the number of interacting neighbors scales as $r_c^3$, this contribution will also vary as $r_c^3u(r_c)\sim r_c^{3-m}$ and will therefore have the same leading-order dependence on $r_c$. Consequently, although shifting modifies the coefficient of the cutoff correction, it does not alter its asymptotic scaling. For the \textsc{Lj} potential, for which the longest-ranged contribution decays as $r^{-6}$, we therefore expect
\[
\Delta\mu(r_c)-\Delta\mu_{\infty}\propto r_c^{-3}.
\]
Consistent with this expectation, Fig.~\ref{fig:Delta-mu-rc}a reveals an excellent linear dependence of $\Delta\mu$ on $r_c^{-3}$. Extrapolation to $r_c^{-3}\rightarrow0$ yields $\Delta\mu_{\infty}/\epsilon=0.28271\pm0.00003$ at $kT/\epsilon=0.5$ and $p\sigma^3/\epsilon=0$. Substitution of this value into Eq.~\eqref{eq:linear-logR-dmusq} gives $\log_{10} R=-7.952\pm0.155$. The rate obtained using $r_c=4\sigma$ therefore differs from the full-potential limit by approximately 1.5 orders of magnitude, whereas those obtained using $r_c\ge 5\sigma$ differ by less than one order of magnitude. Thus, despite the strong sensitivity of nucleation kinetics to the thermodynamic driving force, the rates obtained with cutoffs of $4$--$5\sigma$  already approach the full-potential rate.

We wish to note that the rate obtained here for $r_c=2.5\sigma$ is approximately 0.5 orders of magnitude higher than our earlier estimate of $-18.9453\pm0.0696$ at the same thermodynamic state point and system size.\cite{Hussain2022HowNucleationb} The present calculation employs substantially more extensive sampling of the supercooled liquid basin, with a combined duration of $1.2\times10^6\tau$, compared with $2.5\times10^5\tau$ in the earlier study. As demonstrated previously, the computed nucleation rate can be very sensitive to the duration of basin sampling.\cite{BiJPCB2014} We therefore regard the present value as more reliable. This small difference does not, however, affect any of the aforementioned conclusions.

\subsection{Computational Cost}

\begin{figure*}
 \centering
 \includegraphics[width=0.6419\textwidth]{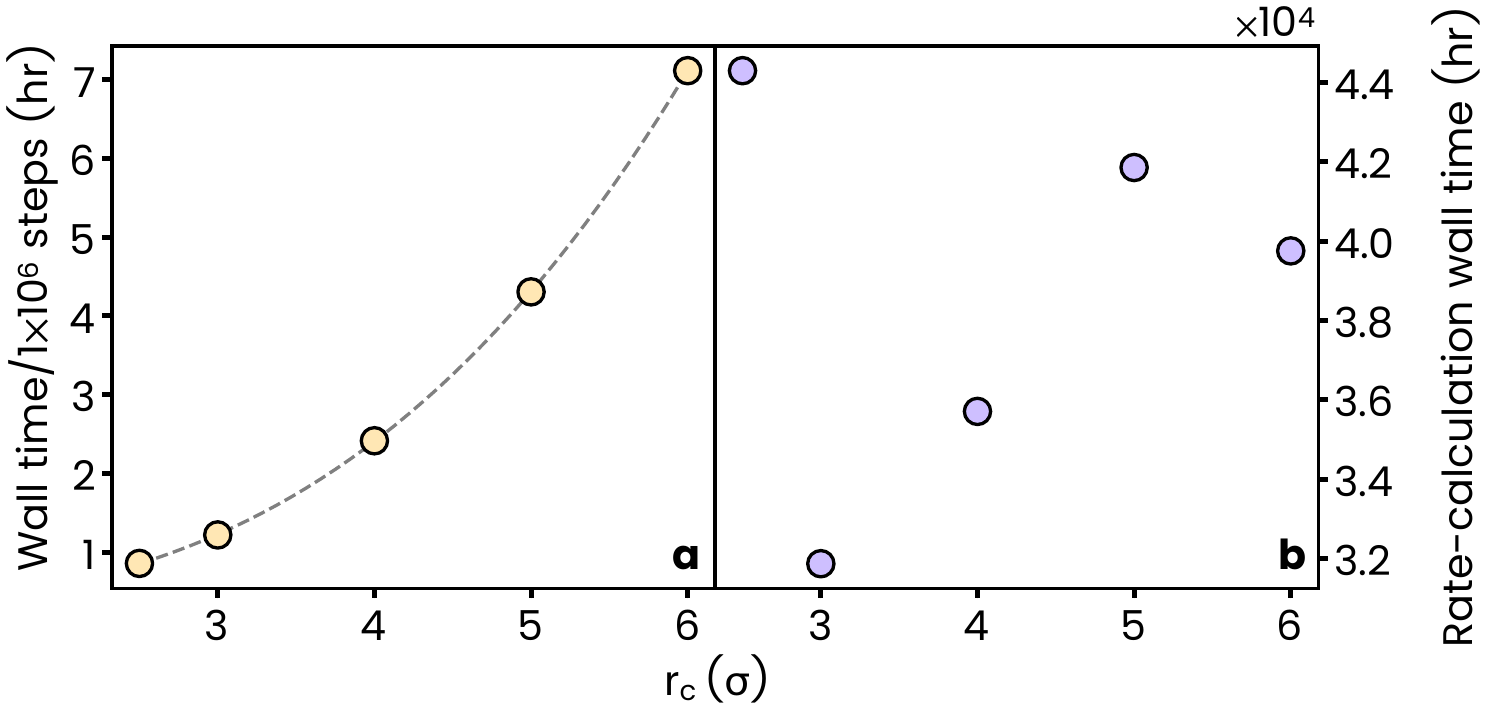}
 \vspace{-10pt}
 \caption{\label{fig:cost-rate} Computational cost as a function of the \textsc{Lj} cutoff radius. (a) Wall time required to perform $10^6$ \textsc{Md} steps on a single CPU core for a system containing 6,912 particles at different cutoff radii. The dashed curve is a cubic fit with $R^2=0.999$. (b) Wall time required to compute the homogeneous nucleation rates reported in Table~\ref{table:LJ_homrates} at different cutoff radii.}
\end{figure*}

\noindent
The choice of \textsc{Lj} cutoff radius in nucleation studies involves a trade-off between accuracy and computational efficiency. Increasing $r_c$ improves the representation of the thermodynamic driving force and nucleation kinetics, but also increases the cost of each \textsc{Md} time step. For a system containing $N_p$ particles at a fixed number density, the number of pairwise interactions scales approximately as $N_p r_c^3$,\cite{BamerArchComputMethodsEng2023} implying a corresponding cubic increase in computational cost. Single-core benchmarks confirm this scaling for the wall-clock time required to complete $10^6$ \textsc{Md}  time steps (Fig.~\ref{fig:cost-rate}a).

\begin{figure*}
 \centering
 \includegraphics[width=0.5431\textwidth]{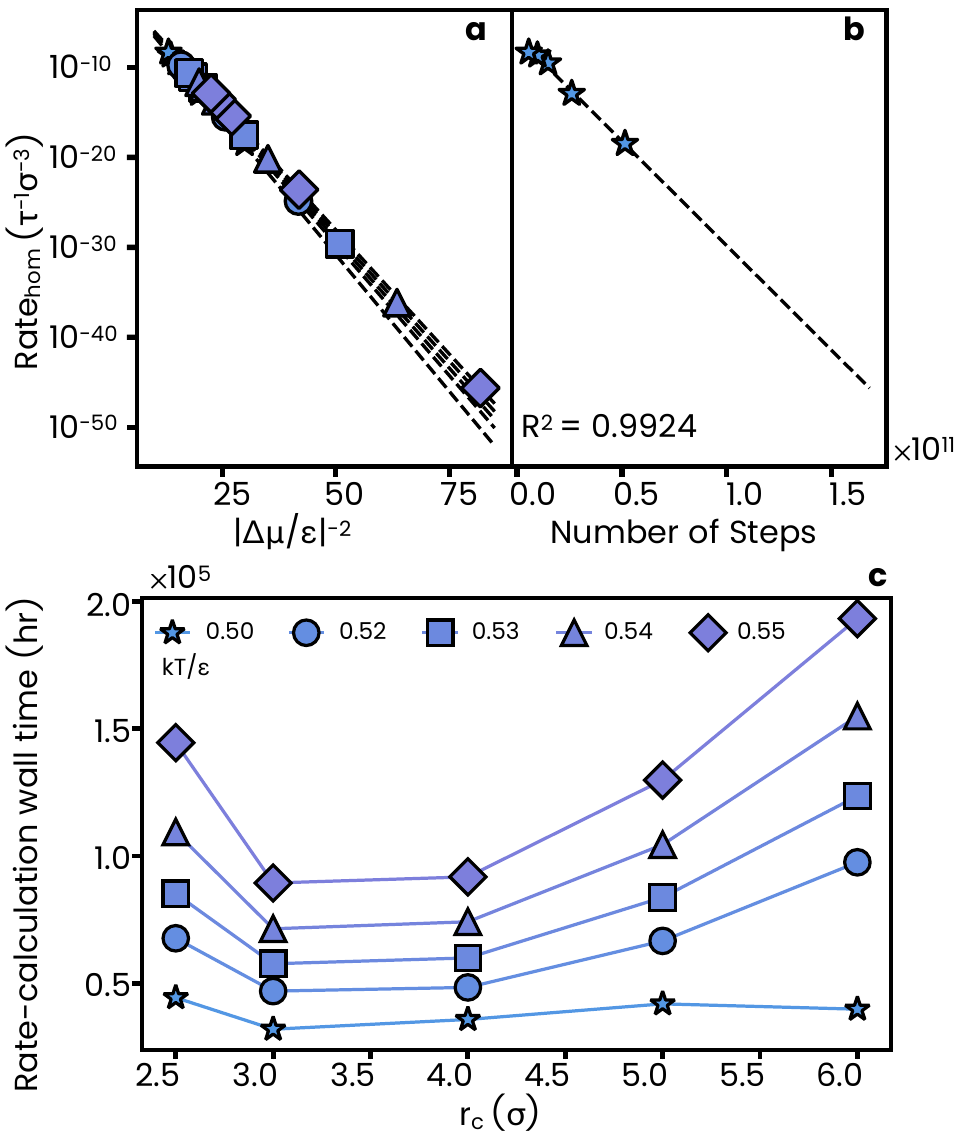}
\vspace{-10pt}
 \caption{\label{fig:cost-allTs} Projected computational cost of estimating homogeneous nucleation rates at different temperatures and cutoff radii. (a) \textsc{Cnt}-based estimates of $R(T,r_c)$ obtained from Eq.~\eqref{eq:rate-other-Ts} using the $\Delta\mu(T,r_c)$ values shown in Fig.~\ref{fig:Delta-mu-rc}a. (b) Linear relationship between $\log_{10} R_{\mathrm{hom}}$ and the number of \textsc{Md} time steps required for the rate calculation ($R^2=0.9924$). (c) Estimated wall times for rate calculations at different $T$ and $r_c$ values, based on the computational benchmarking in Fig.~\ref{fig:cost-rate}a and the linear relationship shown in (b).
}
\vspace{-10pt}
\end{figure*}

The relevant quantity for rate calculations, however, is the total computational cost required to obtain a nucleation rate rather than the cost of an individual \textsc{Md} step. We estimate this cost as
\begin{eqnarray}
\mathcal{C} &=& T_{\text{basin}} + \sum_{i=1}^{N} \left[
N_iT_{s,i} + (M_i-N_i)T_{f,i}
\right],
\end{eqnarray}
where $T_{\text{basin}}$ is the total wall time spent sampling the supercooled liquid basin, $N$ is the number of \textsc{jFfs} iterations, $M_i$ and $N_i$ are the numbers of trial and successful trajectories at milestone $i$, respectively, and $T_{s,i}$ and $T_{f,i}$ are the corresponding average wall times of successful and failing trajectories. All timings are measured using a single processor to avoid complications associated with parallelization efficiency.

As shown in Fig.~\ref{fig:cost-rate}b, $\mathcal{C}$ varies nonmonotonically with $r_c$. Despite having the lowest per-step cost, $r_c=2.5\sigma$ is the most expensive cutoff for the complete nucleation rate calculation. Its smaller thermodynamic driving force results in a substantially lower nucleation rate, requiring an increased number of \textsc{jFfs} milestones and a larger aggregate number of \textsc{Md} steps. Indeed, there is a strong linear relationship between $\log_{10}R_{\mathrm{hom}}$ and the number of required \textsc{Md} time steps ($R^2=0.9924$), as shown in Fig.~\ref{fig:cost-allTs}b. The minimum computational cost therefore occurs at $r_c=3\sigma$, with $r_c=4\sigma$ being the second most efficient choice. Thus, minimizing the cost of individual \textsc{Md} steps does not necessarily minimize the cost of a nucleation rate calculation.

We next examine whether this ordering persists at other temperatures. Using the relationship between nucleation rate and thermodynamic driving force established in Eq.~\eqref{eq:linear-logR-dmusq}, we estimate the cutoff-dependent rates according to
\begin{eqnarray}\label{eq:rate-other-Ts}
\ln R_{\text{hom}}(T,r_c) &=& -\frac{BT_0}{T|\Delta\mu(T,r_c)|^2} + C,
\end{eqnarray}
where $B$ and $C$ are obtained from the rates computed at $kT_0/\epsilon=0.5$. The resulting \textsc{Cnt}-based rate estimates (Fig.~\ref{fig:cost-allTs}a), together with the observed linear relationship between $\log_{10}R_{\text{hom}}$ and the number of required \textsc{Md} time steps at $kT_0/\epsilon=0.5$ (Fig.~\ref{fig:cost-allTs}b),   are used to estimate the projected computational costs depicted in Fig.~\ref{fig:cost-allTs}c. Across the temperature range considered, $r_c=3\sigma$ remains the least expensive choice, while $r_c=4\sigma$ incurs only a modest additional cost.

\begin{figure}
 \centering
 \includegraphics[width=0.4355\textwidth]{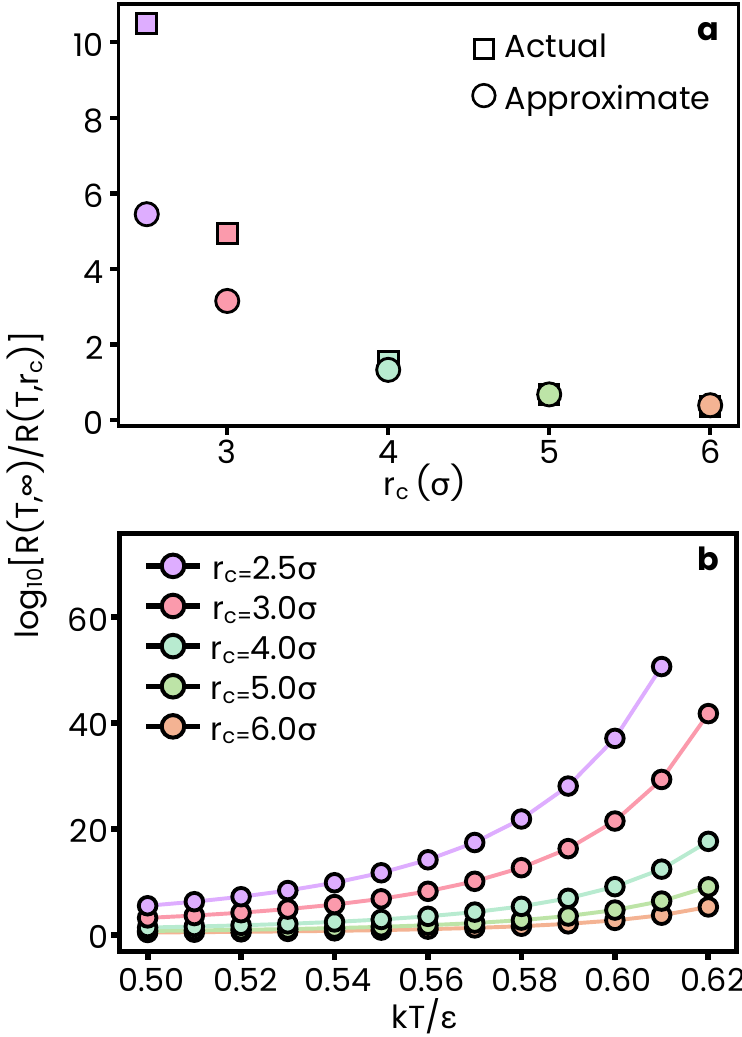}
\caption{\label{fig:logRate-T-rc} (a) Cutoff-induced rate deficit, $\Delta\log_{10} R(T,r_c)=\Delta\ln R(T,r_c)/\ln10$, calculated from the nucleation rates using Eq.~\eqref{eq:rate-deficit-def} (squares), together with the corresponding prediction of Eq.~\eqref{eq:rate-deficit-approx} (circles), at $kT/\epsilon=0.5$. (b) Temperature dependence of the cutoff-induced rate deficit predicted by Eq.~\eqref{eq:rate-deficit-approx} for different $r_c$ values.}
\end{figure}

Computational efficiency alone, however, does not provide an appropriate criterion for cutoff selection. At $kT/\epsilon=0.5$, the rate obtained with $r_c=3\sigma$ differs from the full-potential estimate by nearly five orders of magnitude, whereas the corresponding deviation at $r_c=4\sigma$ is approximately 1.5 orders of magnitude (Fig.~\ref{fig:logRate-T-rc}a). More importantly, the reliability of a finite-cutoff rate can deteriorate with increasing temperature because the nucleation rate becomes progressively more sensitive to small systematic errors in the thermodynamic driving force. A useful cutoff should therefore not only limit the deviation from the full-potential rate, but also lie in a regime where that deviation can be estimated reliably.

To devise a quantitative metric for this criterion, we use the asymptotic relation
\begin{eqnarray}
|\Delta\mu(T,r_c)|
=
|\Delta\mu_{\infty}(T)|
-
\frac{a(T)}{r_c^3}
+\cdots,
\end{eqnarray}
employed in the analysis presented in Fig.~\ref{fig:Delta-mu-rc}a, and make a first-order expansion analogous to that of Vehkam\"aki and Ford.\cite{VehkamkiJCP2000} Defining the positive cutoff-induced rate deficit as
\begin{eqnarray}
\label{eq:rate-deficit-def}
\Delta\ln R(T,r_c)
\equiv
\ln\frac{R(T,\infty)}{R(T,r_c)},
\end{eqnarray}
we obtain
\begin{eqnarray}
\Delta\ln R(T,r_c)
&=&
\frac{BT_0}{T}\left[
\frac{1}{|\Delta\mu(T,r_c)|^2}
-
\frac{1}{|\Delta\mu_{\infty}(T)|^2}
\right]
\notag\\
&\approx&
\frac{2BT_0a(T)}
{T|\Delta\mu_{\infty}(T)|^3r_c^3}.
\label{eq:rate-deficit-approx}
\label{eq:dlnR-approx}
\end{eqnarray}
Here, $a(T)$ is the magnitude of the slope associated with the $r_c^{-3}$ dependence of $|\Delta\mu|$. Equation~\eqref{eq:dlnR-approx} provides an estimate of the finite-cutoff rate deficit using $\Delta\mu_{\infty}$ and rate calculations conducted at finite cutoffs, and explicitly shows how the deficit is amplified as the thermodynamic driving force decreases.

As shown in Fig.~\ref{fig:Delta-mu-rc}b, $a(T)$ is a weak function of temperature over the temperature range considered. Moreover, Figure~\ref{fig:logRate-T-rc}a  demonstrates that the first-order approximation of Eq.~\eqref{eq:dlnR-approx} becomes quantitatively reliable only at sufficiently large $r_c$ values. At $r_c=3\sigma$, for instance, the predicted deficit at $kT/\epsilon=0.5$ is approximately 3.2 orders of magnitude, compared with an actual deficit of nearly five orders of magnitude. At $r_c=4\sigma$, by contrast, the system is much closer to the large-$r_c$ asymptotic regime: the cutoff-induced rate deficit is smaller, and Eq.~\eqref{eq:dlnR-approx}  provides a substantially more accurate estimate of its magnitude.

The sensitivity of the nucleation rate to the cutoff radius becomes increasingly pronounced at smaller supercoolings. As shown in Fig.~\ref{fig:logRate-T-rc}b, the expected rate deficit grows rapidly with increasing temperature because of its inverse cubic dependence on $|\Delta\mu_{\infty}|$ in Eq.~\eqref{eq:dlnR-approx}. At $kT/\epsilon=0.59$, for instance, $r_c=3\sigma$ underestimates the full-potential rate by approximately 17 orders of magnitude, whereas increasing the cutoff to $4\sigma$ reduces the discrepancy to only seven orders of magnitude. The absolute deficit associated with any finite cutoff will ultimately become large sufficiently close to coexistence, where $|\Delta\mu|\rightarrow0$. This regime, however, is also characterized by prohibitively large nucleation barriers and is rarely accessible to direct computational rate calculations. Over the range of supercoolings typically accessible in molecular simulations, $r_c\approx4\sigma$ therefore provides a useful practical cutoff: it retains much of the computational advantage of shorter-ranged potentials while substantially reducing the deviation from the full-potential limit and, importantly, places the system in a regime where the remaining rate deficit can be estimated systematically using Eq.~\eqref{eq:dlnR-approx}.

\subsection{Parallels with Tail Correction Approaches}

\noindent
One might ask whether these truncation-induced errors can be remedied by applying a tail correction that accounts for pair interactions beyond $r_c$. For the mean potential energy per particle, the tail correction is given by\cite{BradleyPhil1932, WoodJCP1957}
\begin{eqnarray}
U_{\text{tail}}(r_c)
&=&
2\pi\rho\int_{r_c}^{\infty} r^2u(r)g(r)\,dr,
\label{eq:tail-general}
\end{eqnarray}
where $u(r)$ is the pair potential and $g(r)$ is the radial distribution function. If the system is spatially homogeneous and $r_c$ is sufficiently large for pair correlations to decay, setting $g(r)\approx1$ for $r\ge r_c$ allows the tail correction in Eq.~\eqref{eq:tail-general} to be evaluated analytically. For the \textsc{Lj} potential, this yields the expression
\begin{eqnarray}
U_{\text{tail}}^{\textsc{LJ}}(r_c)
&=&
\frac{8\pi\rho\epsilon\sigma^3}{3}
\left[
\frac{1}{3}\left(\frac{\sigma}{r_c}\right)^9
-
\left(\frac{\sigma}{r_c}\right)^3
\right].
\label{eq:tail-analytical}
\end{eqnarray}
Although this approximation might be well suited to homogeneous bulk fluids, its applicability to nucleating configurations is less evident. Such configurations simultaneously contain liquid-like, crystalline, and interfacial domains, which cannot generally be represented by a single bulk density and radial distribution function. Moreover, assigning the same mean-field correction to every particle cannot capture variations in the omitted interactions among these structurally distinct domains.

\begin{figure}
    \includegraphics[width=0.4400\textwidth]{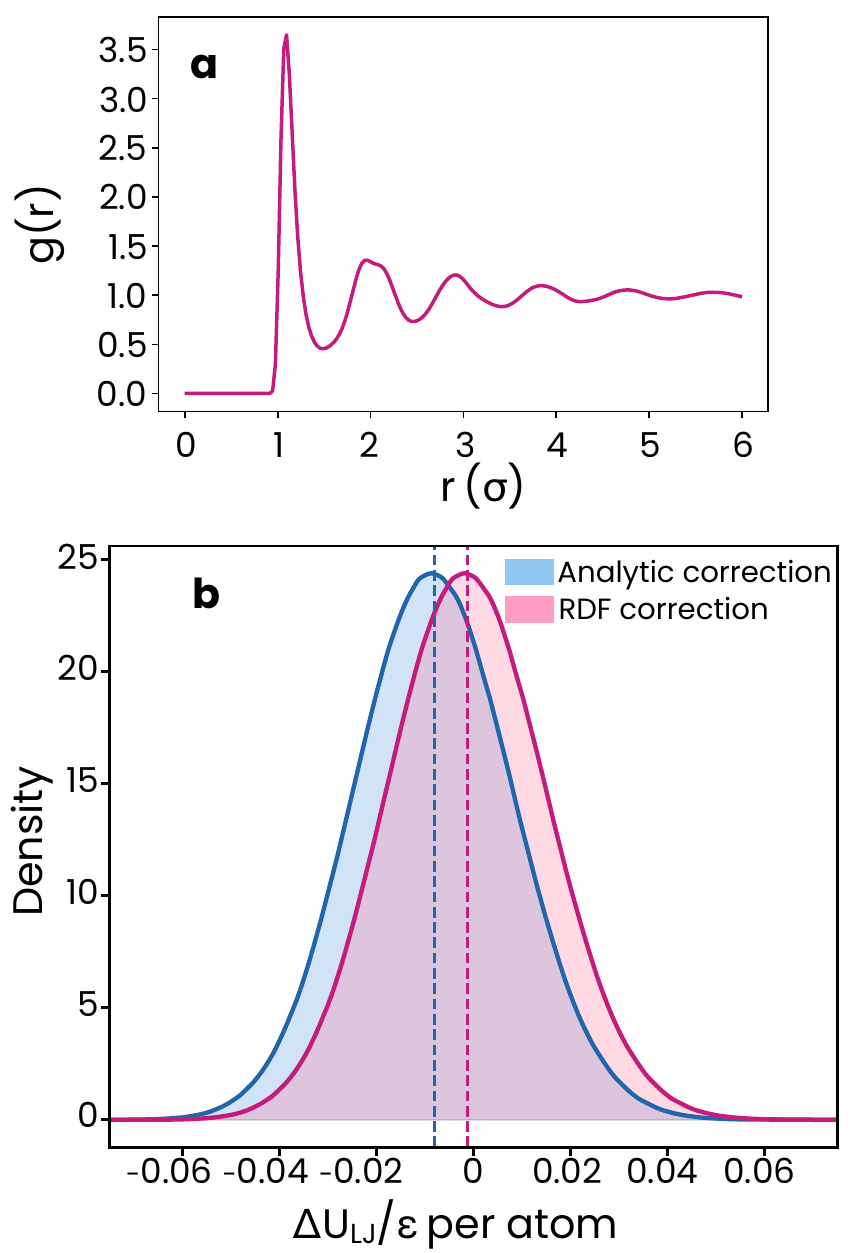}
    \caption{\label{fig:hist}
(a) \textsc{Rdf} of the bulk supercooled liquid at $kT/\epsilon=0.5$ and $p=0$, with $r_c=2.5\sigma$. (b) Statistical distributions of the $\Delta U_{\textsc{LJ}}$'s obtained from Eq.~\eqref{eq:DeltaU-LJ-i} using analytical and \textsc{Rdf}-based tail corrections in the bulk. Dashed vertical lines indicate the mean of each distribution.}
\end{figure}

Before examining tail corrections in nucleating systems, we first assess the extent to which the $g(r)\approx1$ approximation is valid in the homogeneous liquid. We consider supercooled liquid configurations generated with $r_{c,1}=2.5\sigma$ and recompute the potential energy of each particle using the unshifted truncated potential at cutoffs $r_{c,1}=2.5\,\sigma$ and $r_{c,2}=6\,\sigma$.  These values correspond, respectively, to the shortest cutoff considered here and the largest cutoff for which simulation data are available. For each particle, we calculate
\begin{equation}
\Delta U_{\textsc{Lj},i}
=
U_i(r_{c,1})-U_i(r_{c,2})
+
\Delta U_{\text{tail}}(r_{c,1}\rightarrow r_{c,2}),
\label{eq:DeltaU-LJ-i}
\end{equation}
where the predicted contribution from interactions between the two cutoffs is
\begin{eqnarray}
\Delta U_{\text{tail}}(r_{c,1}\rightarrow r_{c,2})
&=&
U_{\text{tail}}(r_{c,1})-U_{\text{tail}}(r_{c,2}).
\end{eqnarray}
An exact tail correction would therefore produce a $\Delta U_{\textsc{Lj}}$ distribution centered at zero.

As shown in Fig.~\ref{fig:hist}b, the analytical correction instead produces a small but systematic shift toward negative values. The omitted interactions are, on average, approximately $0.008\epsilon$ more positive per particle than predicted by Eq.~\eqref{eq:tail-analytical}. Thus, even in the homogeneous supercooled liquid, the $g(r)\approx1$ approximation slightly overestimates the magnitude of the missing attractive interactions at $r_c=2.5\sigma$.

We next evaluate $\Delta U_{\text{tail}}$ by numerically integrating Eq.~\eqref{eq:tail-general} using the \textsc{Rdf} obtained from the \textsc{Md} trajectories (Fig.~\ref{fig:hist}a), as described in Sec.~\ref{subsection:RDFcalc}. This procedure retains the residual pair correlations beyond $2.5\sigma$. The resulting $\Delta U_{\textsc{Lj}}$ distribution is centered much closer to zero, with a mean deviation of only approximately $-10^{-3}\epsilon$ (Fig.~\ref{fig:hist}b). This residual discrepancy is comparable to the uncertainties associated with estimating the \textsc{Rdf} and numerically integrating Eq.~\eqref{eq:tail-general}. The \textsc{Rdf}-based correction therefore accurately recovers the mean omitted pair interactions in the homogeneous liquid, providing a baseline against which its performance in spatially heterogeneous nucleating configurations can be assessed.

\begin{figure}
    \includegraphics[width=0.3194\textwidth]{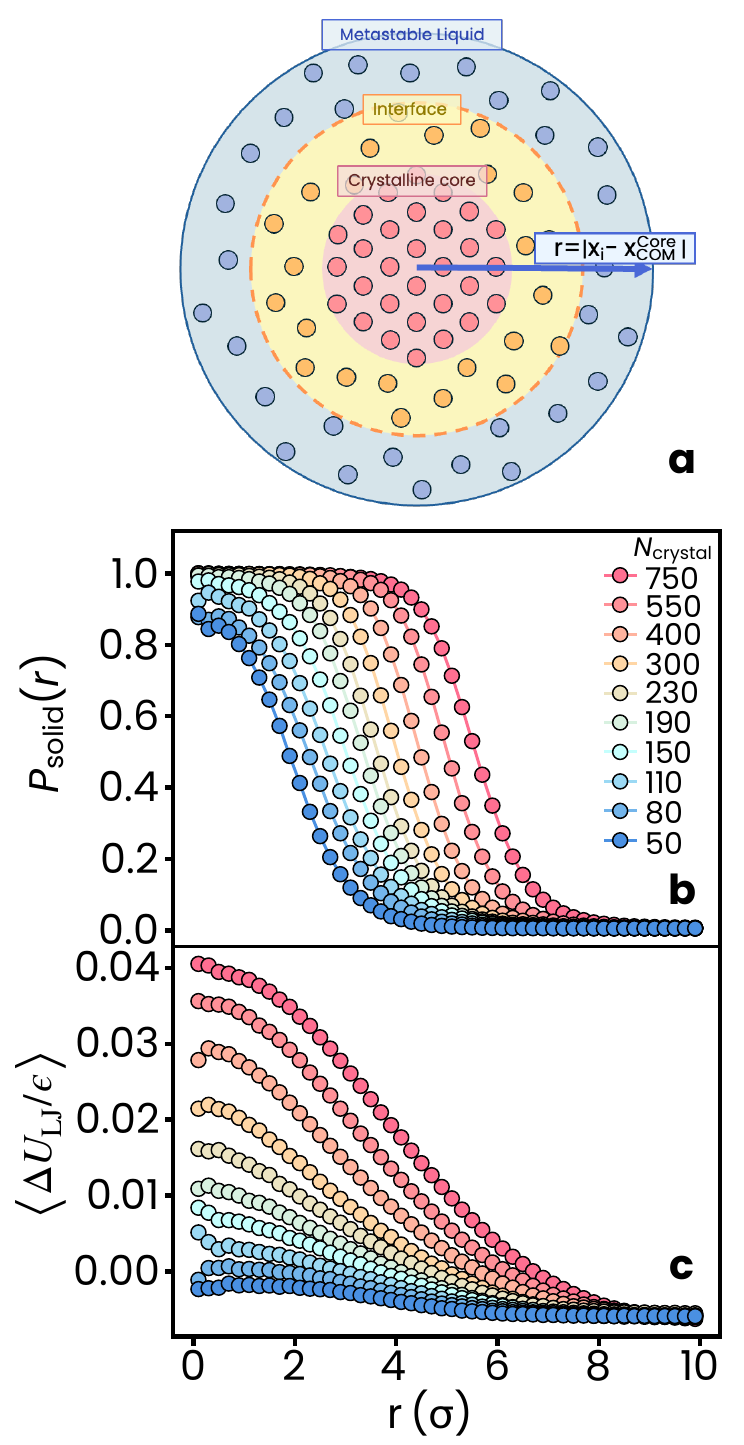}
    \vspace{-10pt}
    \caption{\label{fig:PPR}
(a) Schematic illustration of the radial decomposition of a crystalline nucleus into core (red), interfacial (yellow), and liquid (blue) regions. (b) $P_{\rm solid}(r)$, the probability of observing a solid-like particle at a distance $r$ from the center of mass of the largest crystalline nucleus. (c) $\langle\Delta U_{\textsc{LJ}}\rangle$ as a function of $r$. Colors in (b-c) denote milestones labeled by the size of the largest crystalline cluster, $N_{\rm crystal}$.}
\vspace{-10pt}
\end{figure}

Having established this bulk benchmark, we next assess whether tail corrections remain reliable in nucleating systems, where the assumption of spatial homogeneity is explicitly violated. We focus on crystalline nuclei generated with $r_c=2.5\sigma$, where truncation artifacts are expected to be most pronounced and, because of the smaller driving force, the nuclei are largest. For each configuration, we take the center of mass of the largest crystalline nucleus as the reference point (Fig.~\ref{fig:PPR}a), and compute the radial profile of the fraction of solid-like particles. As shown in Fig.~\ref{fig:PPR}b, the nuclei contain well-defined crystalline cores, with $P_{\text{solid}}(r)\approx1$, separated from the surrounding supercooled liquid, with $P_{\text{solid}}(r)\approx0$, by a narrow interfacial region.

We then repeat the bulk analysis locally by computing $\Delta U_{\textsc{Lj}}$ for each particle using an \textsc{Rdf}-based correction constructed from configurations at each \textsc{jFfs} milestone. Figure~\ref{fig:PPR}c shows the resulting radial average, $\langle\Delta U_{\textsc{Lj}}\rangle$, as a function of the distance $r$ from the center of the nucleus. $\langle\Delta U_{\textsc{Lj}}\rangle$ depends strongly on both $r$ and the size of the crystalline nucleus. It is largest in magnitude near the nucleus core, where the local density and structure differ most strongly from those of the surrounding liquid, and decreases as the environment becomes more liquid-like. However, the residual does not approach zero in the liquid region. Instead, the profiles approach a residual negative offset at large $r$.

This residual offset reflects a basic limitation of applying a homogeneous tail correction to an inhomogeneous nucleating system. The tail correction is evaluated using the global number density of configurations containing both crystalline and disordered domains, which exceeds the local density of the surrounding liquid. As a result, the correction assigns an overly attractive long-range contribution even to particles in bulk liquid-like regions.

These results show that conventional mean-field tail corrections can miss spatially resolved energetic effects that arise during nucleation. This limitation does not apply to all treatments of long-range dispersion.\cite{LagueJPCB2004}  Reciprocal-space approaches, such as dispersion Ewald or \textsc{Lj}-\textsc{Pme},\cite{WennbergJCTC2013, FischerJCTC2015} and generalized corrections for inhomogeneous systems\cite{ChapelaFaradayTrans1977, RowleyJComputPhys1978, LotfiMolSimul1990, BlokhuisMolPhys1995, JanecekJPCB2006} provide alternatives that do not rely on a single bulk density approximation. Such approaches are often computationally more expensive and are not assessed here. For nucleating systems described by finite-range potentials, our results instead suggest that using a shifted potential with a sufficiently large cutoff and extrapolating the resulting rate to the full-potential limit provides a more reliable strategy than applying a conventional homogeneous tail correction at shorter cutoffs.

\section{Implications of Tail Corrections for Phase-Transition Simulations}
\label{section:force-field}

\noindent
The limitations identified here are not confined to the monatomic \textsc{Lj} system, but reflect a broader challenge in applying homogeneous bulk corrections to spatially inhomogeneous systems.\cite{Atherton2022} During nucleation, the parent phase, emerging nucleus, and intervening interface constitute distinct local environments, so a single correction based on the average system density cannot, in general, recover the omitted long-range interactions.

The implications of this observation extend beyond the correction itself to force-field parameterization, as many standard force fields for water\cite{HornJCP2004, AbascalJCP2005, AbascalJCP2005p} and organic liquids\cite{JorgensenJACS1996, Martin1998} are parameterized with prescribed long-range correction schemes, such as analytical tail corrections. In such cases, the correction is effectively part of the force-field definition rather than a separable post-processing choice. Our findings demonstrate that such schemes can become unreliable in inhomogeneous systems, particularly when regions with different densities and structures are irregularly shaped or evolve over time, raising broader concerns about the transferability of such force fields beyond the homogeneous states for which they were parameterized.

As discussed earlier, generalized long-range corrections developed for inhomogeneous systems provide one possible route for addressing this limitation.\cite{ChapelaFaradayTrans1977, RowleyJComputPhys1978, LotfiMolSimul1990, BlokhuisMolPhys1995, JanecekJPCB2006} Most such approaches, however, rely on prescribed geometries or spatial decompositions that remain fixed throughout the simulation. Their extension to nucleation is therefore nontrivial, because the emerging phase forms a finite, irregular, and fluctuating domain whose size and shape evolve along the nucleation pathway.\cite{Sosso2016,Fitzner2017} Accurately treating long-range interactions in such systems will likely require corrections that adapt to the instantaneous spatial organization of the phases rather than imposing a fixed geometry or a single bulk density.

An alternative to real-space tail corrections is to treat long-range dispersion explicitly using reciprocal-space methods, such as Ewald-based treatments\cite{deLeeuwPRSLA1980, LdeeeuwPRSLA1980p} of dispersion interactions.\cite{IntVeldJCP2007, WennbergJCTC2013, FischerJCTC2015} Unlike conventional homogeneous tail corrections, these approaches do not require the neglected interactions to be represented through a single average density and are therefore, in principle, better suited to spatially inhomogeneous systems. Their use, however, changes both the computational cost and the precise definition of the interaction model, and force fields intended for such treatments should be parameterized consistently with the chosen long-range scheme. A systematic comparison of such approaches with large-cutoff finite-range potentials in nucleation calculations would therefore constitute a useful direction for future work.

Our findings therefore suggest that homogeneous tail corrections should generally be avoided in dynamically evolving inhomogeneous systems. A more transferable strategy is to parameterize force fields using sufficiently large, smoothly shifted or switched finite-range potentials, which avoid cutoff discontinuities without imposing a homogeneous mean-field approximation. This approach has longstanding precedent in biomolecular force fields, whose nonbonded interaction parameters were developed for prescribed \textsc{Lj} switching functions rather than analytical tail corrections.\cite{BrooksJComputChem1983, MacKerellJPCB1998, KunzJComputChem2012} Moving forward, force fields intended for heterogeneous applications should be parameterized and validated across both homogeneous and heterogeneous environments, with the cutoff treatment specified as part of the model.

\section{Conclusions}
\label{section:conclusions}
\noindent
In this work, we use \textsc{Md} simulations, \textsc{jFfs}, and free energy calculations to quantify how the cutoff radius in the truncated and shifted \textsc{Lj} potential affects melting thermodynamics, crystal nucleation kinetics, and the computational cost of the rate calculation. Across the range of $r_c$ values considered, the melting temperature changes by approximately 11\%, while the nucleation rate varies by nearly ten orders of magnitude. Classical nucleation theory shows that this kinetic sensitivity is governed primarily by the cutoff dependence of the thermodynamic driving force, with comparatively weak contributions from the interfacial tension and kinetic prefactor. We use this observation not only to estimate the nucleation rate of the full potential, but also to evaluate the cutoff-induced deviation at finite $r_c$'s. The total cost of computing a nucleation rate varies nonmonotonically with $r_c$: despite having the lowest per-step cost, $r_c=2.5\sigma$ yields the most expensive rate calculation because its smaller driving force requires substantially longer sampling. By analyzing the projected scaling of computational cost and rate deviation with temperature and $r_c$, we find  $r_c=4\sigma$ to strike a reasonable balance between computational efficiency and substantially improved agreement with full-potential behavior.

We also examine the reliability of tail corrections and identify two major limitations. Even in a homogeneous liquid, the conventional analytical correction based on the assumption that $g(r)\approx1$ for $r>r_c$ does not fully recover the omitted interactions at short cutoffs, whereas an \textsc{Rdf}-based correction accurately reproduces their mean contribution. In nucleating configurations, however, the problem is more fundamental, as the systematic deviation from the mean-field tail correction becomes position-dependent, varies with nucleus size, and is largest near the crystalline core, where the local density and structure differ most strongly from those of the surrounding liquid. A single correction based on the average properties of the entire system therefore cannot consistently represent the crystalline, interfacial, and liquid regions. For computational studies of nucleation, a shifted potential with a sufficiently large cutoff is consequently preferable to conventional homogeneous tail correction schemes.

A related class of systems in which the treatment of long-range interactions can be consequential comprises those containing charged or polar species. Unlike short-range interactions, which can often be truncated at a finite distance and supplemented with a convergent tail correction, electrostatic interactions between charges and dipoles decay too slowly for such an approach to be generally appropriate. Accurate modeling of these systems therefore requires explicit treatment of long-range electrostatics, commonly through reciprocal-space methods such as Ewald summation.\cite{deLeeuwPRSLA1980, LdeeeuwPRSLA1980p} One subtle aspect of these methods is the treatment of the zero-wavenumber term, which determines the effective electrostatic boundary conditions.\cite{DeLeeuwPhysicaA1981} Although most bulk simulations employ tinfoil, or conducting, boundary conditions,\cite{NeumannJCP1985, EssmannJCP1995} vacuum boundary conditions are more commonly considered in spatially heterogeneous systems.\cite{YehJCP1999} Crystal nucleation in charged or polar systems may not only be subject to particularly strong finite-size effects arising from spurious interactions between a charged or dipolar nucleus and its periodic images, but may also exhibit kinetics, mechanisms, and polymorphic outcomes that depend on the chosen electrostatic boundary conditions. For instance, tinfoil boundary conditions have been shown\cite{LiuJCP2012} to induce the formation of an unphysical ferroelectric ice phase in small simulation cells of ST2 water.\cite{StillingerjCP1974} Surprisingly, relatively little work has systematically examined how electrostatic boundary conditions affect nucleation rates, products, and pathways. Addressing this question therefore represents an important direction for future research.

While this work focuses on the pure \textsc{Lj} system, the physical arguments concerning cutoff sensitivity and tail corrections are more general. Any phase transition involving the emergence of a spatially localized domain with structure or density distinct from the surrounding phase, including crystallization, condensation, liquid-liquid phase separation\cite{Falahati2019ThermodynamicallyBiology} or the formation of ordered mesophases,\cite{XuAnnuRevPhysChem2014} creates an inhomogeneous environment in which bulk density tail corrections may become unreliable. Likewise, because nucleation rates depend exponentially on the thermodynamic driving force, truncation-induced errors in $|\Delta\mu|$ can translate into substantial changes in nucleation rates. Future nucleation studies, particularly those employing \textsc{Lj}-based force fields, should therefore treat the cutoff radius as a convergence parameter that must be explicitly validated.

\begin{acknowledgments}
\noindent 
A.H.-A. gratefully acknowledges the support of the National Science Foundation (\textsc{Nsf}) Grant \textsc{Cbet}-1751971 (\textsc{Career} Award). This work was also supported by the Alfred P. Sloan Foundation.  
F.S.V. acknowledges financial support from the \textsc{Gem} fellowship and the National Science Foundation Graduate Research Fellowship Program (\textsc{Nsf-Grfp}).
These calculations were performed at the Yale Center for Research Computing.  This work used the Extreme Science and Engineering Discovery Environment (\textsc{Xsede}),\cite{TownsComp2014} which is supported by \textsc{Nsf} Grant \textsc{Aci}-1548562. This work used Stampede through allocation \textsc{Chm}240063 from the Advanced Cyberinfrastructure Coordination Ecosystem: Services \& Support (\textsc{Access}) program, which is supported by \textsc{Nsf} Grants \#2138259, \#2138286, \#2138307, \#2137603, and \#2138296.\cite{BoernerACM2023}
\end{acknowledgments}

\bibliographystyle{apsrev4-2}
\bibliography{references_cleaned}

\begin{thebibliography}{116}%
\makeatletter
\providecommand \@ifxundefined [1]{%
 \@ifx{#1\undefined}
}%
\providecommand \@ifnum [1]{%
 \ifnum #1\expandafter \@firstoftwo
 \else \expandafter \@secondoftwo
 \fi
}%
\providecommand \@ifx [1]{%
 \ifx #1\expandafter \@firstoftwo
 \else \expandafter \@secondoftwo
 \fi
}%
\providecommand \natexlab [1]{#1}%
\providecommand \enquote  [1]{``#1''}%
\providecommand \bibnamefont  [1]{#1}%
\providecommand \bibfnamefont [1]{#1}%
\providecommand \citenamefont [1]{#1}%
\providecommand \href@noop [0]{\@secondoftwo}%
\providecommand \href [0]{\begingroup \@sanitize@url \@href}%
\providecommand \@href[1]{\@@startlink{#1}\@@href}%
\providecommand \@@href[1]{\endgroup#1\@@endlink}%
\providecommand \@sanitize@url [0]{\catcode `\\12\catcode `\$12\catcode
  `\&12\catcode `\#12\catcode `\^12\catcode `\_12\catcode `\%12\relax}%
\providecommand \@@startlink[1]{}%
\providecommand \@@endlink[0]{}%
\providecommand \url  [0]{\begingroup\@sanitize@url \@url }%
\providecommand \@url [1]{\endgroup\@href {#1}{\urlprefix }}%
\providecommand \urlprefix  [0]{URL }%
\providecommand \Eprint [0]{\href }%
\providecommand \doibase [0]{https://doi.org/}%
\providecommand \selectlanguage [0]{\@gobble}%
\providecommand \bibinfo  [0]{\@secondoftwo}%
\providecommand \bibfield  [0]{\@secondoftwo}%
\providecommand \translation [1]{[#1]}%
\providecommand \BibitemOpen [0]{}%
\providecommand \bibitemStop [0]{}%
\providecommand \bibitemNoStop [0]{.\EOS\space}%
\providecommand \EOS [0]{\spacefactor3000\relax}%
\providecommand \BibitemShut  [1]{\csname bibitem#1\endcsname}%
\let\auto@bib@innerbib\@empty
\bibitem [{\citenamefont {Sosso}\ \emph
  {et~al.}(2016{\natexlab{a}})\citenamefont {Sosso}, \citenamefont {Chen},
  \citenamefont {Cox}, \citenamefont {Fitzner}, \citenamefont {Pedevilla},
  \citenamefont {Zen},\ and\ \citenamefont {Michaelides}}]{SossoChemRev2016}%
  \BibitemOpen
  \bibfield  {author} {\bibinfo {author} {\bibfnamefont {G.~C.}\ \bibnamefont
  {Sosso}}, \bibinfo {author} {\bibfnamefont {J.}~\bibnamefont {Chen}},
  \bibinfo {author} {\bibfnamefont {S.~J.}\ \bibnamefont {Cox}}, \bibinfo
  {author} {\bibfnamefont {M.}~\bibnamefont {Fitzner}}, \bibinfo {author}
  {\bibfnamefont {P.}~\bibnamefont {Pedevilla}}, \bibinfo {author}
  {\bibfnamefont {A.}~\bibnamefont {Zen}},\ and\ \bibinfo {author}
  {\bibfnamefont {A.}~\bibnamefont {Michaelides}},\ }\href
  {https://doi.org/10.1021/acs.chemrev.5b00744} {\bibfield  {journal} {\bibinfo
   {journal} {Chem. Rev.}\ }\textbf {\bibinfo {volume} {116}},\ \bibinfo
  {pages} {7078} (\bibinfo {year} {2016}{\natexlab{a}})}\BibitemShut {NoStop}%
\bibitem [{\citenamefont {Rein~ten Wolde}\ \emph {et~al.}(1996)\citenamefont
  {Rein~ten Wolde}, \citenamefont {Ruiz-Montero},\ and\ \citenamefont
  {Frenkel}}]{tenWoldeJCP1996}%
  \BibitemOpen
  \bibfield  {author} {\bibinfo {author} {\bibfnamefont {P.}~\bibnamefont
  {Rein~ten Wolde}}, \bibinfo {author} {\bibfnamefont {M.~J.}\ \bibnamefont
  {Ruiz-Montero}},\ and\ \bibinfo {author} {\bibfnamefont {D.}~\bibnamefont
  {Frenkel}},\ }\href {https://doi.org/10.1063/1.471721} {\bibfield  {journal}
  {\bibinfo  {journal} {J. Chem. Phys.}\ }\textbf {\bibinfo {volume} {104}},\
  \bibinfo {pages} {9932} (\bibinfo {year} {1996})}\BibitemShut {NoStop}%
\bibitem [{\citenamefont {Auer}\ and\ \citenamefont
  {Frenkel}(2001)}]{AuerNature2001}%
  \BibitemOpen
  \bibfield  {author} {\bibinfo {author} {\bibfnamefont {S.}~\bibnamefont
  {Auer}}\ and\ \bibinfo {author} {\bibfnamefont {D.}~\bibnamefont {Frenkel}},\
  }\href {https://dx.doi.org/10.1038/35059035} {\bibfield  {journal} {\bibinfo
  {journal} {Nature}\ }\textbf {\bibinfo {volume} {409}},\ \bibinfo {pages}
  {1020} (\bibinfo {year} {2001})}\BibitemShut {NoStop}%
\bibitem [{\citenamefont {Filion}\ \emph {et~al.}(2010)\citenamefont {Filion},
  \citenamefont {Hermes}, \citenamefont {Ni},\ and\ \citenamefont
  {Dijkstra}}]{FilionJCP2010}%
  \BibitemOpen
  \bibfield  {author} {\bibinfo {author} {\bibfnamefont {L.}~\bibnamefont
  {Filion}}, \bibinfo {author} {\bibfnamefont {M.}~\bibnamefont {Hermes}},
  \bibinfo {author} {\bibfnamefont {R.}~\bibnamefont {Ni}},\ and\ \bibinfo
  {author} {\bibfnamefont {M.}~\bibnamefont {Dijkstra}},\ }\href
  {https://doi.org/10.1063/1.3506838} {\bibfield  {journal} {\bibinfo
  {journal} {J. Chem. Phys.}\ }\textbf {\bibinfo {volume} {133}},\ \bibinfo
  {pages} {244115} (\bibinfo {year} {2010})}\BibitemShut {NoStop}%
\bibitem [{\citenamefont {Li}\ \emph {et~al.}(2011)\citenamefont {Li},
  \citenamefont {Donadio}, \citenamefont {Russo},\ and\ \citenamefont
  {Galli}}]{Li2011HomogeneousWater}%
  \BibitemOpen
  \bibfield  {author} {\bibinfo {author} {\bibfnamefont {T.}~\bibnamefont
  {Li}}, \bibinfo {author} {\bibfnamefont {D.}~\bibnamefont {Donadio}},
  \bibinfo {author} {\bibfnamefont {G.}~\bibnamefont {Russo}},\ and\ \bibinfo
  {author} {\bibfnamefont {G.}~\bibnamefont {Galli}},\ }\href
  {https://doi.org/10.1039/C1CP22167A} {\bibfield  {journal} {\bibinfo
  {journal} {Phys. Chem. Chem. Phys.}\ }\textbf {\bibinfo {volume} {13}},\
  \bibinfo {pages} {19807} (\bibinfo {year} {2011})}\BibitemShut {NoStop}%
\bibitem [{\citenamefont {Haji-Akbari}\ \emph {et~al.}(2014)\citenamefont
  {Haji-Akbari}, \citenamefont {DeFever}, \citenamefont {Sarupria},\ and\
  \citenamefont {Debenedetti}}]{HajiAkbariPCCP2014}%
  \BibitemOpen
  \bibfield  {author} {\bibinfo {author} {\bibfnamefont {A.}~\bibnamefont
  {Haji-Akbari}}, \bibinfo {author} {\bibfnamefont {R.~S.}\ \bibnamefont
  {DeFever}}, \bibinfo {author} {\bibfnamefont {S.}~\bibnamefont {Sarupria}},\
  and\ \bibinfo {author} {\bibfnamefont {P.~G.}\ \bibnamefont {Debenedetti}},\
  }\href {https://doi.org/10.1039/c4cp03948c} {\bibfield  {journal} {\bibinfo
  {journal} {Phys. Chem. Chem. Phys.}\ }\textbf {\bibinfo {volume} {16}},\
  \bibinfo {pages} {25916} (\bibinfo {year} {2014})}\BibitemShut {NoStop}%
\bibitem [{\citenamefont {Espinosa}\ \emph {et~al.}(2016)\citenamefont
  {Espinosa}, \citenamefont {Sanz},\ and\ \citenamefont
  {Valeriani}}]{EspinosaJCP2015}%
  \BibitemOpen
  \bibfield  {author} {\bibinfo {author} {\bibfnamefont {J.~R.}\ \bibnamefont
  {Espinosa}}, \bibinfo {author} {\bibfnamefont {E.}~\bibnamefont {Sanz}},\
  and\ \bibinfo {author} {\bibfnamefont {C.}~\bibnamefont {Valeriani}},\ }\href
  {http://dx.doi.org/10.1063/1.4939641} {\bibfield  {journal} {\bibinfo
  {journal} {J. Chem. Phys.}\ }\textbf {\bibinfo {volume} {144}},\ \bibinfo
  {pages} {034501} (\bibinfo {year} {2016})}\BibitemShut {NoStop}%
\bibitem [{\citenamefont {Haji-Akbari}\ and\ \citenamefont
  {Debenedetti}(2015)}]{HajiAkbariPNAS2015}%
  \BibitemOpen
  \bibfield  {author} {\bibinfo {author} {\bibfnamefont {A.}~\bibnamefont
  {Haji-Akbari}}\ and\ \bibinfo {author} {\bibfnamefont {P.~G.}\ \bibnamefont
  {Debenedetti}},\ }\href {https://doi.org/10.1073/pnas.1509267112} {\bibfield
  {journal} {\bibinfo  {journal} {Proc. Natl. Acad. Sci. U.S.A.}\ }\textbf
  {\bibinfo {volume} {112}},\ \bibinfo {pages} {10582} (\bibinfo {year}
  {2015})}\BibitemShut {NoStop}%
\bibitem [{\citenamefont {Haji-Akbari}\ and\ \citenamefont
  {Debenedetti}(2017)}]{HajiAkbariPNAS2017}%
  \BibitemOpen
  \bibfield  {author} {\bibinfo {author} {\bibfnamefont {A.}~\bibnamefont
  {Haji-Akbari}}\ and\ \bibinfo {author} {\bibfnamefont {P.~G.}\ \bibnamefont
  {Debenedetti}},\ }\href {https://doi.org/10.1073/pnas.1620999114} {\bibfield
  {journal} {\bibinfo  {journal} {Proc. Natl. Acad. Sci. U.S.A.}\ }\textbf
  {\bibinfo {volume} {114}},\ \bibinfo {pages} {3316} (\bibinfo {year}
  {2017})}\BibitemShut {NoStop}%
\bibitem [{\citenamefont {Jiang}\ \emph
  {et~al.}(2018{\natexlab{a}})\citenamefont {Jiang}, \citenamefont
  {Haji-Akbari}, \citenamefont {Debenedetti},\ and\ \citenamefont
  {Panagiotopoulos}}]{JiangJCP2018p}%
  \BibitemOpen
  \bibfield  {author} {\bibinfo {author} {\bibfnamefont {H.}~\bibnamefont
  {Jiang}}, \bibinfo {author} {\bibfnamefont {A.}~\bibnamefont {Haji-Akbari}},
  \bibinfo {author} {\bibfnamefont {P.~G.}\ \bibnamefont {Debenedetti}},\ and\
  \bibinfo {author} {\bibfnamefont {A.~Z.}\ \bibnamefont {Panagiotopoulos}},\
  }\href {https://doi.org/10.1063/1.5016554} {\bibfield  {journal} {\bibinfo
  {journal} {J. Chem. Phys.}\ }\textbf {\bibinfo {volume} {148}},\ \bibinfo
  {pages} {044505} (\bibinfo {year} {2018}{\natexlab{a}})}\BibitemShut
  {NoStop}%
\bibitem [{\citenamefont {Yuan}\ \emph {et~al.}(2023)\citenamefont {Yuan},
  \citenamefont {DeFever}, \citenamefont {Zhou}, \citenamefont
  {Cortes-Morales},\ and\ \citenamefont {Sarupria}}]{YuanJPCB2023}%
  \BibitemOpen
  \bibfield  {author} {\bibinfo {author} {\bibfnamefont {T.}~\bibnamefont
  {Yuan}}, \bibinfo {author} {\bibfnamefont {R.~S.}\ \bibnamefont {DeFever}},
  \bibinfo {author} {\bibfnamefont {J.}~\bibnamefont {Zhou}}, \bibinfo {author}
  {\bibfnamefont {E.~C.}\ \bibnamefont {Cortes-Morales}},\ and\ \bibinfo
  {author} {\bibfnamefont {S.}~\bibnamefont {Sarupria}},\ }\href
  {https://doi.org/10.1021/acs.jpcb.3c00910} {\bibfield  {journal} {\bibinfo
  {journal} {J. Phys. Chem. B}\ }\textbf {\bibinfo {volume} {127}},\ \bibinfo
  {pages} {4112} (\bibinfo {year} {2023})}\BibitemShut {NoStop}%
\bibitem [{\citenamefont {Mendelev}\ \emph {et~al.}(2008)\citenamefont
  {Mendelev}, \citenamefont {Kramer}, \citenamefont {Becker},\ and\
  \citenamefont {Asta}}]{MendelevPhilosMag2008}%
  \BibitemOpen
  \bibfield  {author} {\bibinfo {author} {\bibfnamefont {M.}~\bibnamefont
  {Mendelev}}, \bibinfo {author} {\bibfnamefont {M.}~\bibnamefont {Kramer}},
  \bibinfo {author} {\bibfnamefont {C.~A.}\ \bibnamefont {Becker}},\ and\
  \bibinfo {author} {\bibfnamefont {M.}~\bibnamefont {Asta}},\ }\href
  {https://doi.org/10.1080/14786430802206482} {\bibfield  {journal} {\bibinfo
  {journal} {Philos. Mag.}\ }\textbf {\bibinfo {volume} {88}},\ \bibinfo
  {pages} {1723} (\bibinfo {year} {2008})}\BibitemShut {NoStop}%
\bibitem [{\citenamefont {Espinosa}\ \emph {et~al.}(2014)\citenamefont
  {Espinosa}, \citenamefont {Sanz}, \citenamefont {Valeriani},\ and\
  \citenamefont {Vega}}]{EspinosaJCP2014}%
  \BibitemOpen
  \bibfield  {author} {\bibinfo {author} {\bibfnamefont {J.}~\bibnamefont
  {Espinosa}}, \bibinfo {author} {\bibfnamefont {E.}~\bibnamefont {Sanz}},
  \bibinfo {author} {\bibfnamefont {C.}~\bibnamefont {Valeriani}},\ and\
  \bibinfo {author} {\bibfnamefont {C.}~\bibnamefont {Vega}},\ }\href
  {https://doi.org/10.1063/1.4897524} {\bibfield  {journal} {\bibinfo
  {journal} {J. Chem. Phys.}\ }\textbf {\bibinfo {volume} {141}},\ \bibinfo
  {pages} {18C529} (\bibinfo {year} {2014})}\BibitemShut {NoStop}%
\bibitem [{\citenamefont {Gianetti}\ \emph {et~al.}(2016)\citenamefont
  {Gianetti}, \citenamefont {Haji-Akbari}, \citenamefont {Longinotti},\ and\
  \citenamefont {Debenedetti}}]{GianettiPCCP2016}%
  \BibitemOpen
  \bibfield  {author} {\bibinfo {author} {\bibfnamefont {M.~M.}\ \bibnamefont
  {Gianetti}}, \bibinfo {author} {\bibfnamefont {A.}~\bibnamefont
  {Haji-Akbari}}, \bibinfo {author} {\bibfnamefont {M.~P.}\ \bibnamefont
  {Longinotti}},\ and\ \bibinfo {author} {\bibfnamefont {P.~G.}\ \bibnamefont
  {Debenedetti}},\ }\href {https://doi.org/10.1039/C5CP06535F} {\bibfield
  {journal} {\bibinfo  {journal} {Phys. Chem. Chem. Phys.}\ }\textbf {\bibinfo
  {volume} {18}},\ \bibinfo {pages} {4102} (\bibinfo {year}
  {2016})}\BibitemShut {NoStop}%
\bibitem [{\citenamefont {Jiang}\ \emph
  {et~al.}(2018{\natexlab{b}})\citenamefont {Jiang}, \citenamefont
  {Debenedetti},\ and\ \citenamefont {Panagiotopoulos}}]{JiangJCP2018}%
  \BibitemOpen
  \bibfield  {author} {\bibinfo {author} {\bibfnamefont {H.}~\bibnamefont
  {Jiang}}, \bibinfo {author} {\bibfnamefont {P.~G.}\ \bibnamefont
  {Debenedetti}},\ and\ \bibinfo {author} {\bibfnamefont {A.~Z.}\ \bibnamefont
  {Panagiotopoulos}},\ }\href {https://doi.org/10.1063/1.5053652} {\bibfield
  {journal} {\bibinfo  {journal} {J. Chem. Phys.}\ }\textbf {\bibinfo {volume}
  {149}},\ \bibinfo {pages} {141102} (\bibinfo {year}
  {2018}{\natexlab{b}})}\BibitemShut {NoStop}%
\bibitem [{\citenamefont {Sanchez-Burgos}\ \emph {et~al.}(2022)\citenamefont
  {Sanchez-Burgos}, \citenamefont {Tejedor}, \citenamefont {Vega},
  \citenamefont {Conde}, \citenamefont {Sanz}, \citenamefont {Ramirez},\ and\
  \citenamefont {Espinosa}}]{SanchezJCP2022}%
  \BibitemOpen
  \bibfield  {author} {\bibinfo {author} {\bibfnamefont {I.}~\bibnamefont
  {Sanchez-Burgos}}, \bibinfo {author} {\bibfnamefont {A.~R.}\ \bibnamefont
  {Tejedor}}, \bibinfo {author} {\bibfnamefont {C.}~\bibnamefont {Vega}},
  \bibinfo {author} {\bibfnamefont {M.~M.}\ \bibnamefont {Conde}}, \bibinfo
  {author} {\bibfnamefont {E.}~\bibnamefont {Sanz}}, \bibinfo {author}
  {\bibfnamefont {J.}~\bibnamefont {Ramirez}},\ and\ \bibinfo {author}
  {\bibfnamefont {J.~R.}\ \bibnamefont {Espinosa}},\ }\href
  {https://doi.org/10.1063/5.0101383} {\bibfield  {journal} {\bibinfo
  {journal} {J. Chem. Phys.}\ }\textbf {\bibinfo {volume} {157}},\ \bibinfo
  {pages} {094503} (\bibinfo {year} {2022})}\BibitemShut {NoStop}%
\bibitem [{\citenamefont {Lauricella}\ \emph {et~al.}(2023)\citenamefont
  {Lauricella}, \citenamefont {Meloni},\ and\ \citenamefont
  {Ciccotti}}]{LauricellaJCP2023}%
  \BibitemOpen
  \bibfield  {author} {\bibinfo {author} {\bibfnamefont {M.}~\bibnamefont
  {Lauricella}}, \bibinfo {author} {\bibfnamefont {S.}~\bibnamefont {Meloni}},\
  and\ \bibinfo {author} {\bibfnamefont {G.}~\bibnamefont {Ciccotti}},\ }\href
  {https://doi.org/10.1063/5.0140951} {\bibfield  {journal} {\bibinfo
  {journal} {J. Chem. Phys.}\ }\textbf {\bibinfo {volume} {158}},\ \bibinfo
  {pages} {164501} (\bibinfo {year} {2023})}\BibitemShut {NoStop}%
\bibitem [{\citenamefont {Minh}\ \emph {et~al.}(2025)\citenamefont {Minh},
  \citenamefont {Hall}, \citenamefont {DeFever},\ and\ \citenamefont
  {Sarupria}}]{MinhJPCB2025}%
  \BibitemOpen
  \bibfield  {author} {\bibinfo {author} {\bibfnamefont {P.}~\bibnamefont
  {Minh}}, \bibinfo {author} {\bibfnamefont {S.~W.}\ \bibnamefont {Hall}},
  \bibinfo {author} {\bibfnamefont {R.~S.}\ \bibnamefont {DeFever}},\ and\
  \bibinfo {author} {\bibfnamefont {S.}~\bibnamefont {Sarupria}},\ }\href
  {https://doi.org/10.1021/acs.jpcb.5c04269} {\bibfield  {journal} {\bibinfo
  {journal} {J. Phys. Chem. B}\ }\textbf {\bibinfo {volume} {129}},\ \bibinfo
  {pages} {8976} (\bibinfo {year} {2025})}\BibitemShut {NoStop}%
\bibitem [{\citenamefont {Sandberg}\ \emph {et~al.}(2025)\citenamefont
  {Sandberg}, \citenamefont {Voigtmann}, \citenamefont {Devijver},\ and\
  \citenamefont {Jakse}}]{SandbergJCP2025}%
  \BibitemOpen
  \bibfield  {author} {\bibinfo {author} {\bibfnamefont {J.}~\bibnamefont
  {Sandberg}}, \bibinfo {author} {\bibfnamefont {T.}~\bibnamefont {Voigtmann}},
  \bibinfo {author} {\bibfnamefont {E.}~\bibnamefont {Devijver}},\ and\
  \bibinfo {author} {\bibfnamefont {N.}~\bibnamefont {Jakse}},\ }\href
  {https://doi.org/10.1063/5.0299431} {\bibfield  {journal} {\bibinfo
  {journal} {J. Chem. Phys.}\ }\textbf {\bibinfo {volume} {163}},\ \bibinfo
  {pages} {214503} (\bibinfo {year} {2025})}\BibitemShut {NoStop}%
\bibitem [{\citenamefont {Hussain}\ and\ \citenamefont
  {Haji-Akbari}(2021)}]{Hussain2021HowNucleation}%
  \BibitemOpen
  \bibfield  {author} {\bibinfo {author} {\bibfnamefont {S.}~\bibnamefont
  {Hussain}}\ and\ \bibinfo {author} {\bibfnamefont {A.}~\bibnamefont
  {Haji-Akbari}},\ }\href {https://doi.org/10.1063/5.0026355} {\bibfield
  {journal} {\bibinfo  {journal} {J. Chem. Phys.}\ }\textbf {\bibinfo {volume}
  {154}},\ \bibinfo {pages} {014108} (\bibinfo {year} {2021})}\BibitemShut
  {NoStop}%
\bibitem [{\citenamefont {Hussain}\ and\ \citenamefont
  {Haji-Akbari}(2022)}]{Hussain2022HowNucleationb}%
  \BibitemOpen
  \bibfield  {author} {\bibinfo {author} {\bibfnamefont {S.}~\bibnamefont
  {Hussain}}\ and\ \bibinfo {author} {\bibfnamefont {A.}~\bibnamefont
  {Haji-Akbari}},\ }\href {https://doi.org/10.1063/5.0079702} {\bibfield
  {journal} {\bibinfo  {journal} {J. Chem. Phys.}\ }\textbf {\bibinfo {volume}
  {156}},\ \bibinfo {pages} {054503} (\bibinfo {year} {2022})}\BibitemShut
  {NoStop}%
\bibitem [{\citenamefont {Karmakar}\ \emph {et~al.}(2019)\citenamefont
  {Karmakar}, \citenamefont {Piaggi},\ and\ \citenamefont
  {Parrinello}}]{KarmakarJCTC2019}%
  \BibitemOpen
  \bibfield  {author} {\bibinfo {author} {\bibfnamefont {T.}~\bibnamefont
  {Karmakar}}, \bibinfo {author} {\bibfnamefont {P.~M.}\ \bibnamefont
  {Piaggi}},\ and\ \bibinfo {author} {\bibfnamefont {M.}~\bibnamefont
  {Parrinello}},\ }\href {https://doi.org/10.1021/acs.jctc.9b00795} {\bibfield
  {journal} {\bibinfo  {journal} {J. Chem. Theory Comput.}\ }\textbf {\bibinfo
  {volume} {15}},\ \bibinfo {pages} {6923} (\bibinfo {year}
  {2019})}\BibitemShut {NoStop}%
\bibitem [{\citenamefont {Montero De~Hijes}\ \emph {et~al.}(2020)\citenamefont
  {Montero De~Hijes}, \citenamefont {Espinosa}, \citenamefont {Bianco},
  \citenamefont {Sanz},\ and\ \citenamefont {Vega}}]{MonteroJPC2020}%
  \BibitemOpen
  \bibfield  {author} {\bibinfo {author} {\bibfnamefont {P.}~\bibnamefont
  {Montero De~Hijes}}, \bibinfo {author} {\bibfnamefont {J.~R.}\ \bibnamefont
  {Espinosa}}, \bibinfo {author} {\bibfnamefont {V.}~\bibnamefont {Bianco}},
  \bibinfo {author} {\bibfnamefont {E.}~\bibnamefont {Sanz}},\ and\ \bibinfo
  {author} {\bibfnamefont {C.}~\bibnamefont {Vega}},\ }\href
  {https://doi.org/10.1021/acs.jpcc.0c00816} {\bibfield  {journal} {\bibinfo
  {journal} {J. Phys. Chem. C}\ }\textbf {\bibinfo {volume} {124}},\ \bibinfo
  {pages} {8795} (\bibinfo {year} {2020})}\BibitemShut {NoStop}%
\bibitem [{\citenamefont {Fiorucci}\ \emph {et~al.}(2020)\citenamefont
  {Fiorucci}, \citenamefont {Coli}, \citenamefont {Padding},\ and\
  \citenamefont {Dijkstra}}]{FiorucciJCP2020}%
  \BibitemOpen
  \bibfield  {author} {\bibinfo {author} {\bibfnamefont {G.}~\bibnamefont
  {Fiorucci}}, \bibinfo {author} {\bibfnamefont {G.~M.}\ \bibnamefont {Coli}},
  \bibinfo {author} {\bibfnamefont {J.~T.}\ \bibnamefont {Padding}},\ and\
  \bibinfo {author} {\bibfnamefont {M.}~\bibnamefont {Dijkstra}},\ }\href
  {https://doi.org/10.1063/1.5137815} {\bibfield  {journal} {\bibinfo
  {journal} {J. Chem. Phys.}\ }\textbf {\bibinfo {volume} {152}},\ \bibinfo
  {pages} {064903} (\bibinfo {year} {2020})}\BibitemShut {NoStop}%
\bibitem [{\citenamefont {Blow}\ \emph {et~al.}(2023)\citenamefont {Blow},
  \citenamefont {Tribello}, \citenamefont {Sosso},\ and\ \citenamefont
  {Quigley}}]{BlowJCP2023}%
  \BibitemOpen
  \bibfield  {author} {\bibinfo {author} {\bibfnamefont {K.~E.}\ \bibnamefont
  {Blow}}, \bibinfo {author} {\bibfnamefont {G.~A.}\ \bibnamefont {Tribello}},
  \bibinfo {author} {\bibfnamefont {G.~C.}\ \bibnamefont {Sosso}},\ and\
  \bibinfo {author} {\bibfnamefont {D.}~\bibnamefont {Quigley}},\ }\href
  {https://doi.org/10.1063/5.0152343} {\bibfield  {journal} {\bibinfo
  {journal} {J. Chem. Phys.}\ }\textbf {\bibinfo {volume} {158}},\ \bibinfo
  {pages} {224102} (\bibinfo {year} {2023})}\BibitemShut {NoStop}%
\bibitem [{\citenamefont {Gispen}\ \emph {et~al.}(2024)\citenamefont {Gispen},
  \citenamefont {Espinosa}, \citenamefont {Sanz}, \citenamefont {Vega},\ and\
  \citenamefont {Dijkstra}}]{GispenJCP2024}%
  \BibitemOpen
  \bibfield  {author} {\bibinfo {author} {\bibfnamefont {W.}~\bibnamefont
  {Gispen}}, \bibinfo {author} {\bibfnamefont {J.~R.}\ \bibnamefont
  {Espinosa}}, \bibinfo {author} {\bibfnamefont {E.}~\bibnamefont {Sanz}},
  \bibinfo {author} {\bibfnamefont {C.}~\bibnamefont {Vega}},\ and\ \bibinfo
  {author} {\bibfnamefont {M.}~\bibnamefont {Dijkstra}},\ }\href
  {https://doi.org/10.1063/5.0204540} {\bibfield  {journal} {\bibinfo
  {journal} {J. Chem. Phys.}\ }\textbf {\bibinfo {volume} {160}},\ \bibinfo
  {pages} {174501} (\bibinfo {year} {2024})}\BibitemShut {NoStop}%
\bibitem [{\citenamefont {Domingues}\ \emph {et~al.}(2024)\citenamefont
  {Domingues}, \citenamefont {Hussain},\ and\ \citenamefont
  {Haji-Akbari}}]{Domingues2024DivergenceCrystals}%
  \BibitemOpen
  \bibfield  {author} {\bibinfo {author} {\bibfnamefont {T.~S.}\ \bibnamefont
  {Domingues}}, \bibinfo {author} {\bibfnamefont {S.}~\bibnamefont {Hussain}},\
  and\ \bibinfo {author} {\bibfnamefont {A.}~\bibnamefont {Haji-Akbari}},\
  }\href {https://doi.org/10.1021/acs.jpclett.3c03561} {\bibfield  {journal}
  {\bibinfo  {journal} {J. Phys. Chem. Lett.}\ }\textbf {\bibinfo {volume}
  {15}},\ \bibinfo {pages} {1279} (\bibinfo {year} {2024})}\BibitemShut
  {NoStop}%
\bibitem [{\citenamefont {Sinaeian}\ and\ \citenamefont
  {Haji-Akbari}(2025)}]{Sinaeian2025}%
  \BibitemOpen
  \bibfield  {author} {\bibinfo {author} {\bibfnamefont {K.}~\bibnamefont
  {Sinaeian}}\ and\ \bibinfo {author} {\bibfnamefont {A.}~\bibnamefont
  {Haji-Akbari}},\ }\href {https://doi.org/10.1063/5.0263587} {\bibfield
  {journal} {\bibinfo  {journal} {J. Chem. Phys.}\ }\textbf {\bibinfo {volume}
  {162}},\ \bibinfo {pages} {164102} (\bibinfo {year} {2025})}\BibitemShut
  {NoStop}%
\bibitem [{\citenamefont {Carr{\'e}}\ \emph {et~al.}(2007)\citenamefont
  {Carr{\'e}}, \citenamefont {Berthier}, \citenamefont {Horbach}, \citenamefont
  {Ispas},\ and\ \citenamefont {Kob}}]{CareJCP2007}%
  \BibitemOpen
  \bibfield  {author} {\bibinfo {author} {\bibfnamefont {A.}~\bibnamefont
  {Carr{\'e}}}, \bibinfo {author} {\bibfnamefont {L.}~\bibnamefont {Berthier}},
  \bibinfo {author} {\bibfnamefont {J.}~\bibnamefont {Horbach}}, \bibinfo
  {author} {\bibfnamefont {S.}~\bibnamefont {Ispas}},\ and\ \bibinfo {author}
  {\bibfnamefont {W.}~\bibnamefont {Kob}},\ }\href
  {https://doi.org/10.1063/1.2777136} {\bibfield  {journal} {\bibinfo
  {journal} {J. Chem. Phys.}\ }\textbf {\bibinfo {volume} {127}},\ \bibinfo
  {pages} {114512} (\bibinfo {year} {2007})}\BibitemShut {NoStop}%
\bibitem [{\citenamefont {Ingebrigtsen}\ \emph {et~al.}(2012)\citenamefont
  {Ingebrigtsen}, \citenamefont {Schr{\o}der},\ and\ \citenamefont
  {Dyre}}]{IngebrigtsenPRX2012}%
  \BibitemOpen
  \bibfield  {author} {\bibinfo {author} {\bibfnamefont {T.~S.}\ \bibnamefont
  {Ingebrigtsen}}, \bibinfo {author} {\bibfnamefont {T.~B.}\ \bibnamefont
  {Schr{\o}der}},\ and\ \bibinfo {author} {\bibfnamefont {J.~C.}\ \bibnamefont
  {Dyre}},\ }\href {https://doi.org/10.1103/PhysRevX.2.011011} {\bibfield
  {journal} {\bibinfo  {journal} {Phys. Rev. X}\ }\textbf {\bibinfo {volume}
  {2}},\ \bibinfo {pages} {011011} (\bibinfo {year} {2012})}\BibitemShut
  {NoStop}%
\bibitem [{\citenamefont {Jones}(1924)}]{Jones1924OnTemperature}%
  \BibitemOpen
  \bibfield  {author} {\bibinfo {author} {\bibfnamefont {J.~E.}\ \bibnamefont
  {Jones}},\ }\href {https://doi.org/10.1098/RSPA.1924.0081} {\bibfield
  {journal} {\bibinfo  {journal} {Proc. R. Soc. Lond. A}\ }\textbf {\bibinfo
  {volume} {106}},\ \bibinfo {pages} {441} (\bibinfo {year}
  {1924})}\BibitemShut {NoStop}%
\bibitem [{\citenamefont {Rosenbluth}\ and\ \citenamefont
  {Rosenbluth}(1954)}]{RosenbluthJCP1954}%
  \BibitemOpen
  \bibfield  {author} {\bibinfo {author} {\bibfnamefont {M.~N.}\ \bibnamefont
  {Rosenbluth}}\ and\ \bibinfo {author} {\bibfnamefont {A.~W.}\ \bibnamefont
  {Rosenbluth}},\ }\href {https://doi.org/10.1063/1.1740207} {\bibfield
  {journal} {\bibinfo  {journal} {J. Chem. Phys.}\ }\textbf {\bibinfo {volume}
  {22}},\ \bibinfo {pages} {881} (\bibinfo {year} {1954})}\BibitemShut
  {NoStop}%
\bibitem [{\citenamefont {Weeks}\ \emph {et~al.}(1971)\citenamefont {Weeks},
  \citenamefont {Chandler},\ and\ \citenamefont
  {Andersen}}]{Weeks1971RoleLiquids}%
  \BibitemOpen
  \bibfield  {author} {\bibinfo {author} {\bibfnamefont {J.~D.}\ \bibnamefont
  {Weeks}}, \bibinfo {author} {\bibfnamefont {D.}~\bibnamefont {Chandler}},\
  and\ \bibinfo {author} {\bibfnamefont {H.~C.}\ \bibnamefont {Andersen}},\
  }\href {https://doi.org/10.1063/1.1674820} {\bibfield  {journal} {\bibinfo
  {journal} {J. Chem. Phys.}\ }\textbf {\bibinfo {volume} {54}},\ \bibinfo
  {pages} {5237} (\bibinfo {year} {1971})}\BibitemShut {NoStop}%
\bibitem [{\citenamefont {Stoddard}\ and\ \citenamefont
  {Ford}(1973)}]{StoddardPRA1973}%
  \BibitemOpen
  \bibfield  {author} {\bibinfo {author} {\bibfnamefont {S.~D.}\ \bibnamefont
  {Stoddard}}\ and\ \bibinfo {author} {\bibfnamefont {J.}~\bibnamefont
  {Ford}},\ }\href {https://doi.org/10.1103/PhysRevA.8.1504} {\bibfield
  {journal} {\bibinfo  {journal} {Phys. Rev. A}\ }\textbf {\bibinfo {volume}
  {8}},\ \bibinfo {pages} {1504} (\bibinfo {year} {1973})}\BibitemShut
  {NoStop}%
\bibitem [{\citenamefont {Brooks}\ \emph {et~al.}(1983)\citenamefont {Brooks},
  \citenamefont {Bruccoleri}, \citenamefont {Olafson}, \citenamefont {States},
  \citenamefont {Swaminathan},\ and\ \citenamefont
  {Karplus}}]{BrooksJComputChem1983}%
  \BibitemOpen
  \bibfield  {author} {\bibinfo {author} {\bibfnamefont {B.~R.}\ \bibnamefont
  {Brooks}}, \bibinfo {author} {\bibfnamefont {R.~E.}\ \bibnamefont
  {Bruccoleri}}, \bibinfo {author} {\bibfnamefont {B.~D.}\ \bibnamefont
  {Olafson}}, \bibinfo {author} {\bibfnamefont {D.~J.}\ \bibnamefont {States}},
  \bibinfo {author} {\bibfnamefont {S.~A.}\ \bibnamefont {Swaminathan}},\ and\
  \bibinfo {author} {\bibfnamefont {M.}~\bibnamefont {Karplus}},\ }\href
  {https://doi.org/10.1002/jcc.540040211} {\bibfield  {journal} {\bibinfo
  {journal} {J. Comput. Chem.}\ }\textbf {\bibinfo {volume} {4}},\ \bibinfo
  {pages} {187} (\bibinfo {year} {1983})}\BibitemShut {NoStop}%
\bibitem [{\citenamefont {Steinbach}\ and\ \citenamefont
  {Brooks}(1994)}]{SteinbachJComputChem1994}%
  \BibitemOpen
  \bibfield  {author} {\bibinfo {author} {\bibfnamefont {P.~J.}\ \bibnamefont
  {Steinbach}}\ and\ \bibinfo {author} {\bibfnamefont {B.~R.}\ \bibnamefont
  {Brooks}},\ }\href {https://doi.org/10.1002/jcc.540150702} {\bibfield
  {journal} {\bibinfo  {journal} {J. Comput. Chem.}\ }\textbf {\bibinfo
  {volume} {15}},\ \bibinfo {pages} {667} (\bibinfo {year} {1994})}\BibitemShut
  {NoStop}%
\bibitem [{\citenamefont {Wood}\ and\ \citenamefont
  {Parker}(1957)}]{WoodJCP1957}%
  \BibitemOpen
  \bibfield  {author} {\bibinfo {author} {\bibfnamefont {W.~W.}\ \bibnamefont
  {Wood}}\ and\ \bibinfo {author} {\bibfnamefont {F.~R.}\ \bibnamefont
  {Parker}},\ }\href {https://doi.org/10.1063/1.1743822} {\bibfield  {journal}
  {\bibinfo  {journal} {J. Chem. Phys.}\ }\textbf {\bibinfo {volume} {27}},\
  \bibinfo {pages} {720} (\bibinfo {year} {1957})}\BibitemShut {NoStop}%
\bibitem [{\citenamefont {Toxvaerd}\ and\ \citenamefont
  {Dyre}(2011)}]{ToxvaerdJCP2011}%
  \BibitemOpen
  \bibfield  {author} {\bibinfo {author} {\bibfnamefont {S.}~\bibnamefont
  {Toxvaerd}}\ and\ \bibinfo {author} {\bibfnamefont {J.~C.}\ \bibnamefont
  {Dyre}},\ }\href {https://doi.org/10.1063/1.3558787} {\bibfield  {journal}
  {\bibinfo  {journal} {J. Chem. Phys.}\ }\textbf {\bibinfo {volume} {134}},\
  \bibinfo {pages} {081102} (\bibinfo {year} {2011})}\BibitemShut {NoStop}%
\bibitem [{\citenamefont {Lautenschlaeger}\ and\ \citenamefont
  {Hasse}(2019)}]{LautenschlaegerFluidPhaseEquil2019}%
  \BibitemOpen
  \bibfield  {author} {\bibinfo {author} {\bibfnamefont {M.~P.}\ \bibnamefont
  {Lautenschlaeger}}\ and\ \bibinfo {author} {\bibfnamefont {H.}~\bibnamefont
  {Hasse}},\ }\href {https://dx.doi.org/10.1016/j.fluid.2018.10.019} {\bibfield
   {journal} {\bibinfo  {journal} {Fluid Phase Equilib.}\ }\textbf {\bibinfo
  {volume} {482}},\ \bibinfo {pages} {38} (\bibinfo {year} {2019})}\BibitemShut
  {NoStop}%
\bibitem [{\citenamefont {Shaul}\ \emph {et~al.}(2010)\citenamefont {Shaul},
  \citenamefont {Schultz},\ and\ \citenamefont {Kofke}}]{ShaulCCCC2010}%
  \BibitemOpen
  \bibfield  {author} {\bibinfo {author} {\bibfnamefont {K.~R.}\ \bibnamefont
  {Shaul}}, \bibinfo {author} {\bibfnamefont {A.~J.}\ \bibnamefont {Schultz}},\
  and\ \bibinfo {author} {\bibfnamefont {D.~A.}\ \bibnamefont {Kofke}},\ }\href
  {https://doi.org/10.1135/cccc2009113} {\bibfield  {journal} {\bibinfo
  {journal} {Collect. Czech. Chem. Commun.}\ }\textbf {\bibinfo {volume}
  {75}},\ \bibinfo {pages} {447} (\bibinfo {year} {2010})}\BibitemShut
  {NoStop}%
\bibitem [{\citenamefont {Smit}(1992)}]{Smit1992PhaseFluids}%
  \BibitemOpen
  \bibfield  {author} {\bibinfo {author} {\bibfnamefont {B.}~\bibnamefont
  {Smit}},\ }\href {https://doi.org/10.1063/1.462271} {\bibfield  {journal}
  {\bibinfo  {journal} {J. Chem. Phys.}\ }\textbf {\bibinfo {volume} {96}},\
  \bibinfo {pages} {8639} (\bibinfo {year} {1992})}\BibitemShut {NoStop}%
\bibitem [{\citenamefont {Trokhymchuk}\ and\ \citenamefont
  {Alejandre}(1999)}]{Trokhymchuk1999ComputerAnswers}%
  \BibitemOpen
  \bibfield  {author} {\bibinfo {author} {\bibfnamefont {A.}~\bibnamefont
  {Trokhymchuk}}\ and\ \bibinfo {author} {\bibfnamefont {J.}~\bibnamefont
  {Alejandre}},\ }\href {https://doi.org/10.1063/1.480192} {\bibfield
  {journal} {\bibinfo  {journal} {J. Chem. Phys.}\ }\textbf {\bibinfo {volume}
  {111}},\ \bibinfo {pages} {8510} (\bibinfo {year} {1999})}\BibitemShut
  {NoStop}%
\bibitem [{\citenamefont {Vrabec}\ \emph {et~al.}(2006)\citenamefont {Vrabec},
  \citenamefont {Kedia}, \citenamefont {Fuchs},\ and\ \citenamefont
  {Hasse}}]{VrabecMolPhys2006}%
  \BibitemOpen
  \bibfield  {author} {\bibinfo {author} {\bibfnamefont {J.}~\bibnamefont
  {Vrabec}}, \bibinfo {author} {\bibfnamefont {G.~K.}\ \bibnamefont {Kedia}},
  \bibinfo {author} {\bibfnamefont {G.}~\bibnamefont {Fuchs}},\ and\ \bibinfo
  {author} {\bibfnamefont {H.}~\bibnamefont {Hasse}},\ }\href
  {https://doi.org/10.1080/00268970600556774} {\bibfield  {journal} {\bibinfo
  {journal} {Mol. Phys.}\ }\textbf {\bibinfo {volume} {104}},\ \bibinfo {pages}
  {1509} (\bibinfo {year} {2006})}\BibitemShut {NoStop}%
\bibitem [{\citenamefont {Wang}\ \emph {et~al.}(2024)\citenamefont {Wang},
  \citenamefont {Ram{\'\i}rez-Hinestrosa}, \citenamefont {Dobnikar},\ and\
  \citenamefont {Frenkel}}]{WangPCCP2024}%
  \BibitemOpen
  \bibfield  {author} {\bibinfo {author} {\bibfnamefont {X.}~\bibnamefont
  {Wang}}, \bibinfo {author} {\bibfnamefont {S.}~\bibnamefont
  {Ram{\'\i}rez-Hinestrosa}}, \bibinfo {author} {\bibfnamefont
  {J.}~\bibnamefont {Dobnikar}},\ and\ \bibinfo {author} {\bibfnamefont
  {D.}~\bibnamefont {Frenkel}},\ }\href {https://doi.org/10.1039/d4cp90118e}
  {\bibfield  {journal} {\bibinfo  {journal} {Phys. Chem. Chem. Phys.}\
  }\textbf {\bibinfo {volume} {26}},\ \bibinfo {pages} {24714} (\bibinfo {year}
  {2024})}\BibitemShut {NoStop}%
\bibitem [{\citenamefont {Moro}\ \emph {et~al.}(2024)\citenamefont {Moro},
  \citenamefont {Ballenegger}, \citenamefont {Underwood},\ and\ \citenamefont
  {Wilding}}]{MoroPCCP2024}%
  \BibitemOpen
  \bibfield  {author} {\bibinfo {author} {\bibfnamefont {O.~S.}\ \bibnamefont
  {Moro}}, \bibinfo {author} {\bibfnamefont {V.}~\bibnamefont {Ballenegger}},
  \bibinfo {author} {\bibfnamefont {T.~L.}\ \bibnamefont {Underwood}},\ and\
  \bibinfo {author} {\bibfnamefont {N.~B.}\ \bibnamefont {Wilding}},\ }\href
  {https://doi.org/10.1039/d3cp05474h} {\bibfield  {journal} {\bibinfo
  {journal} {Phys. Chem. Chem. Phys.}\ }\textbf {\bibinfo {volume} {26}},\
  \bibinfo {pages} {7573} (\bibinfo {year} {2024})}\BibitemShut {NoStop}%
\bibitem [{\citenamefont {Nijmeijer}\ \emph {et~al.}(1988)\citenamefont
  {Nijmeijer}, \citenamefont {Bakker}, \citenamefont {Bruin},\ and\
  \citenamefont {Sikkenk}}]{NijmeijerJCP1988}%
  \BibitemOpen
  \bibfield  {author} {\bibinfo {author} {\bibfnamefont {M.}~\bibnamefont
  {Nijmeijer}}, \bibinfo {author} {\bibfnamefont {A.}~\bibnamefont {Bakker}},
  \bibinfo {author} {\bibfnamefont {C.}~\bibnamefont {Bruin}},\ and\ \bibinfo
  {author} {\bibfnamefont {J.}~\bibnamefont {Sikkenk}},\ }\href
  {https://dx.doi.org/10.1063/1.454902} {\bibfield  {journal} {\bibinfo
  {journal} {J. Chem. Phys.}\ }\textbf {\bibinfo {volume} {89}},\ \bibinfo
  {pages} {3789} (\bibinfo {year} {1988})}\BibitemShut {NoStop}%
\bibitem [{\citenamefont {Shen}\ \emph {et~al.}(2007)\citenamefont {Shen},
  \citenamefont {Mountain},\ and\ \citenamefont {Errington}}]{ShenJPCB2007}%
  \BibitemOpen
  \bibfield  {author} {\bibinfo {author} {\bibfnamefont {V.~K.}\ \bibnamefont
  {Shen}}, \bibinfo {author} {\bibfnamefont {R.~D.}\ \bibnamefont {Mountain}},\
  and\ \bibinfo {author} {\bibfnamefont {J.~R.}\ \bibnamefont {Errington}},\
  }\href {https://doi.org/10.1021/jp070374f} {\bibfield  {journal} {\bibinfo
  {journal} {J. Phys. Chem. B}\ }\textbf {\bibinfo {volume} {111}},\ \bibinfo
  {pages} {6198} (\bibinfo {year} {2007})}\BibitemShut {NoStop}%
\bibitem [{\citenamefont {Takahashi}\ \emph {et~al.}(2007)\citenamefont
  {Takahashi}, \citenamefont {Yasuoka},\ and\ \citenamefont
  {Narumi}}]{TakahashiJCP2007}%
  \BibitemOpen
  \bibfield  {author} {\bibinfo {author} {\bibfnamefont {K.}~\bibnamefont
  {Takahashi}}, \bibinfo {author} {\bibfnamefont {K.}~\bibnamefont {Yasuoka}},\
  and\ \bibinfo {author} {\bibfnamefont {T.}~\bibnamefont {Narumi}},\ }\href
  {https://doi.org/10.1063/1.2775929} {\bibfield  {journal} {\bibinfo
  {journal} {J. Chem. Phys.}\ }\textbf {\bibinfo {volume} {127}},\ \bibinfo
  {pages} {114511} (\bibinfo {year} {2007})}\BibitemShut {NoStop}%
\bibitem [{\citenamefont {Bugel}\ and\ \citenamefont
  {Galliero}(2008)}]{BugelChemPhys2008}%
  \BibitemOpen
  \bibfield  {author} {\bibinfo {author} {\bibfnamefont {M.}~\bibnamefont
  {Bugel}}\ and\ \bibinfo {author} {\bibfnamefont {G.}~\bibnamefont
  {Galliero}},\ }\href {https://dx.doi.org/10.1016/j.chemphys.2008.06.013}
  {\bibfield  {journal} {\bibinfo  {journal} {Chem. Phys.}\ }\textbf {\bibinfo
  {volume} {352}},\ \bibinfo {pages} {249} (\bibinfo {year}
  {2008})}\BibitemShut {NoStop}%
\bibitem [{\citenamefont {Santra}\ \emph {et~al.}(2008)\citenamefont {Santra},
  \citenamefont {Chakrabarty},\ and\ \citenamefont
  {Bagchi}}]{Santra2008Gas-liquidSystem}%
  \BibitemOpen
  \bibfield  {author} {\bibinfo {author} {\bibfnamefont {M.}~\bibnamefont
  {Santra}}, \bibinfo {author} {\bibfnamefont {S.}~\bibnamefont
  {Chakrabarty}},\ and\ \bibinfo {author} {\bibfnamefont {B.}~\bibnamefont
  {Bagchi}},\ }\href {https://doi.org/10.1063/1.3037241} {\bibfield  {journal}
  {\bibinfo  {journal} {J. Chem. Phys.}\ }\textbf {\bibinfo {volume} {129}},\
  \bibinfo {pages} {234704} (\bibinfo {year} {2008})}\BibitemShut {NoStop}%
\bibitem [{\citenamefont {Wang}\ \emph {et~al.}(2020)\citenamefont {Wang},
  \citenamefont {Ram{\'\i}rez-Hinestrosa}, \citenamefont {Dobnikar},\ and\
  \citenamefont {Frenkel}}]{WangPCCP2020}%
  \BibitemOpen
  \bibfield  {author} {\bibinfo {author} {\bibfnamefont {X.}~\bibnamefont
  {Wang}}, \bibinfo {author} {\bibfnamefont {S.}~\bibnamefont
  {Ram{\'\i}rez-Hinestrosa}}, \bibinfo {author} {\bibfnamefont
  {J.}~\bibnamefont {Dobnikar}},\ and\ \bibinfo {author} {\bibfnamefont
  {D.}~\bibnamefont {Frenkel}},\ }\href {https://doi.org/10.1039/c9cp05445f}
  {\bibfield  {journal} {\bibinfo  {journal} {Phys. Chem. Chem. Phys.}\
  }\textbf {\bibinfo {volume} {22}},\ \bibinfo {pages} {10624} (\bibinfo {year}
  {2020})}\BibitemShut {NoStop}%
\bibitem [{\citenamefont {Mastny}\ and\ \citenamefont
  {de~Pablo}(2007)}]{Mastny2007MeltingPotential}%
  \BibitemOpen
  \bibfield  {author} {\bibinfo {author} {\bibfnamefont {E.~A.}\ \bibnamefont
  {Mastny}}\ and\ \bibinfo {author} {\bibfnamefont {J.~J.}\ \bibnamefont
  {de~Pablo}},\ }\href {https://doi.org/10.1063/1.2753149} {\bibfield
  {journal} {\bibinfo  {journal} {J. Chem. Phys.}\ }\textbf {\bibinfo {volume}
  {127}},\ \bibinfo {pages} {104504} (\bibinfo {year} {2007})}\BibitemShut
  {NoStop}%
\bibitem [{\citenamefont {Morris}\ and\ \citenamefont
  {Song}(2002)}]{MorrisJCP2002}%
  \BibitemOpen
  \bibfield  {author} {\bibinfo {author} {\bibfnamefont {J.~R.}\ \bibnamefont
  {Morris}}\ and\ \bibinfo {author} {\bibfnamefont {X.}~\bibnamefont {Song}},\
  }\href {https://doi.org/10.1063/1.1474581} {\bibfield  {journal} {\bibinfo
  {journal} {J. Chem. Phys.}\ }\textbf {\bibinfo {volume} {116}},\ \bibinfo
  {pages} {9352} (\bibinfo {year} {2002})}\BibitemShut {NoStop}%
\bibitem [{\citenamefont {Ahmed}\ and\ \citenamefont
  {Sadus}(2010)}]{Ahmed2010EffectFluids}%
  \BibitemOpen
  \bibfield  {author} {\bibinfo {author} {\bibfnamefont {A.}~\bibnamefont
  {Ahmed}}\ and\ \bibinfo {author} {\bibfnamefont {R.~J.}\ \bibnamefont
  {Sadus}},\ }\href {https://doi.org/10.1063/1.3481102} {\bibfield  {journal}
  {\bibinfo  {journal} {J. Chem. Phys.}\ }\textbf {\bibinfo {volume} {133}},\
  \bibinfo {pages} {124515} (\bibinfo {year} {2010})}\BibitemShut {NoStop}%
\bibitem [{\citenamefont {Schultz}\ and\ \citenamefont
  {Kofke}(2018)}]{SchultzJCP2018}%
  \BibitemOpen
  \bibfield  {author} {\bibinfo {author} {\bibfnamefont {A.~J.}\ \bibnamefont
  {Schultz}}\ and\ \bibinfo {author} {\bibfnamefont {D.~A.}\ \bibnamefont
  {Kofke}},\ }\href {https://doi.org/10.1063/1.5053714} {\bibfield  {journal}
  {\bibinfo  {journal} {J. Chem. Phys.}\ }\textbf {\bibinfo {volume} {149}},\
  \bibinfo {pages} {204508} (\bibinfo {year} {2018})}\BibitemShut {NoStop}%
\bibitem [{\citenamefont {Volmer}\ and\ \citenamefont
  {Weber}(1926)}]{Volmer1926KeimbildungGebilden}%
  \BibitemOpen
  \bibfield  {author} {\bibinfo {author} {\bibfnamefont {M.}~\bibnamefont
  {Volmer}}\ and\ \bibinfo {author} {\bibfnamefont {A.}~\bibnamefont {Weber}},\
  }\href {https://doi.org/10.1515/ZPCH-1926-11927} {\bibfield  {journal}
  {\bibinfo  {journal} {Z. Phys. Chem.}\ }\textbf {\bibinfo {volume} {119U}},\
  \bibinfo {pages} {277} (\bibinfo {year} {1926})}\BibitemShut {NoStop}%
\bibitem [{\citenamefont {Becker}\ and\ \citenamefont
  {D\"{o}ring}(1935)}]{Becker1935KinetischeDampfen}%
  \BibitemOpen
  \bibfield  {author} {\bibinfo {author} {\bibfnamefont {R.}~\bibnamefont
  {Becker}}\ and\ \bibinfo {author} {\bibfnamefont {W.}~\bibnamefont
  {D\"{o}ring}},\ }\href {https://dx.doi.org/10.1002/ANDP.19354160806}
  {\bibfield  {journal} {\bibinfo  {journal} {Ann. Phys.}\ }\textbf {\bibinfo
  {volume} {416}},\ \bibinfo {pages} {719} (\bibinfo {year}
  {1935})}\BibitemShut {NoStop}%
\bibitem [{\citenamefont {Turnbull}\ and\ \citenamefont
  {Fisher}(1949)}]{Turnbull1949RateSystems}%
  \BibitemOpen
  \bibfield  {author} {\bibinfo {author} {\bibfnamefont {D.}~\bibnamefont
  {Turnbull}}\ and\ \bibinfo {author} {\bibfnamefont {J.~C.}\ \bibnamefont
  {Fisher}},\ }\href {https://doi.org/10.1063/1.1747055} {\bibfield  {journal}
  {\bibinfo  {journal} {J. Chem. Phys.}\ }\textbf {\bibinfo {volume} {17}},\
  \bibinfo {pages} {71} (\bibinfo {year} {1949})}\BibitemShut {NoStop}%
\bibitem [{\citenamefont {Chapela}\ \emph {et~al.}(1977)\citenamefont
  {Chapela}, \citenamefont {Saville}, \citenamefont {Thompson},\ and\
  \citenamefont {Rowlinson}}]{ChapelaFaradayTrans1977}%
  \BibitemOpen
  \bibfield  {author} {\bibinfo {author} {\bibfnamefont {G.~A.}\ \bibnamefont
  {Chapela}}, \bibinfo {author} {\bibfnamefont {G.}~\bibnamefont {Saville}},
  \bibinfo {author} {\bibfnamefont {S.~M.}\ \bibnamefont {Thompson}},\ and\
  \bibinfo {author} {\bibfnamefont {J.~S.}\ \bibnamefont {Rowlinson}},\ }\href
  {https://doi.org/10.1039/F29777301133} {\bibfield  {journal} {\bibinfo
  {journal} {J. Chem. Soc., Faraday Trans. 2}\ }\textbf {\bibinfo {volume}
  {73}},\ \bibinfo {pages} {1133} (\bibinfo {year} {1977})}\BibitemShut
  {NoStop}%
\bibitem [{\citenamefont {Lotfi}\ \emph {et~al.}(1990)\citenamefont {Lotfi},
  \citenamefont {Vrabec},\ and\ \citenamefont {Fischer}}]{LotfiMolSimul1990}%
  \BibitemOpen
  \bibfield  {author} {\bibinfo {author} {\bibfnamefont {A.}~\bibnamefont
  {Lotfi}}, \bibinfo {author} {\bibfnamefont {J.}~\bibnamefont {Vrabec}},\ and\
  \bibinfo {author} {\bibfnamefont {J.}~\bibnamefont {Fischer}},\ }\href
  {https://doi.org/10.1080/08927029008022133} {\bibfield  {journal} {\bibinfo
  {journal} {Mol. Simul.}\ }\textbf {\bibinfo {volume} {5}},\ \bibinfo {pages}
  {233} (\bibinfo {year} {1990})}\BibitemShut {NoStop}%
\bibitem [{\citenamefont {Blokhuis}\ \emph {et~al.}(1995)\citenamefont
  {Blokhuis}, \citenamefont {Bedeaux}, \citenamefont {Holcomb},\ and\
  \citenamefont {Zollweg}}]{BlokhuisMolPhys1995}%
  \BibitemOpen
  \bibfield  {author} {\bibinfo {author} {\bibfnamefont {E.}~\bibnamefont
  {Blokhuis}}, \bibinfo {author} {\bibfnamefont {D.}~\bibnamefont {Bedeaux}},
  \bibinfo {author} {\bibfnamefont {C.}~\bibnamefont {Holcomb}},\ and\ \bibinfo
  {author} {\bibfnamefont {J.}~\bibnamefont {Zollweg}},\ }\href
  {https://dx.doi.org/10.1080/00268979500101371} {\bibfield  {journal}
  {\bibinfo  {journal} {Mol. Phys.}\ }\textbf {\bibinfo {volume} {85}},\
  \bibinfo {pages} {665} (\bibinfo {year} {1995})}\BibitemShut {NoStop}%
\bibitem [{\citenamefont {Jane\v{c}ek}(2006)}]{JanecekJPCB2006}%
  \BibitemOpen
  \bibfield  {author} {\bibinfo {author} {\bibfnamefont {J.}~\bibnamefont
  {Jane\v{c}ek}},\ }\href {https://doi.org/10.1021/jp056344z} {\bibfield
  {journal} {\bibinfo  {journal} {J. Phys. Chem. B}\ }\textbf {\bibinfo
  {volume} {110}},\ \bibinfo {pages} {6264} (\bibinfo {year}
  {2006})}\BibitemShut {NoStop}%
\bibitem [{\citenamefont {Rowley}\ \emph {et~al.}(1978)\citenamefont {Rowley},
  \citenamefont {Nicholson},\ and\ \citenamefont
  {Parsonage}}]{RowleyJComputPhys1978}%
  \BibitemOpen
  \bibfield  {author} {\bibinfo {author} {\bibfnamefont {L.}~\bibnamefont
  {Rowley}}, \bibinfo {author} {\bibfnamefont {D.}~\bibnamefont {Nicholson}},\
  and\ \bibinfo {author} {\bibfnamefont {N.}~\bibnamefont {Parsonage}},\ }\href
  {https://dx.doi.org/10.1016/0021-9991(78)90099-2} {\bibfield  {journal}
  {\bibinfo  {journal} {J. Comput. Phys.}\ }\textbf {\bibinfo {volume} {26}},\
  \bibinfo {pages} {66} (\bibinfo {year} {1978})}\BibitemShut {NoStop}%
\bibitem [{\citenamefont {Jablonka}\ \emph {et~al.}(2019)\citenamefont
  {Jablonka}, \citenamefont {Ongari},\ and\ \citenamefont
  {Smit}}]{Jablonka2019}%
  \BibitemOpen
  \bibfield  {author} {\bibinfo {author} {\bibfnamefont {K.~M.}\ \bibnamefont
  {Jablonka}}, \bibinfo {author} {\bibfnamefont {D.}~\bibnamefont {Ongari}},\
  and\ \bibinfo {author} {\bibfnamefont {B.}~\bibnamefont {Smit}},\ }\href
  {https://doi.org/10.1021/acs.jctc.9b00586} {\bibfield  {journal} {\bibinfo
  {journal} {J. Chem. Theory Comput.}\ }\textbf {\bibinfo {volume} {15}},\
  \bibinfo {pages} {5635} (\bibinfo {year} {2019})}\BibitemShut {NoStop}%
\bibitem [{\citenamefont
  {Haji-Akbari}(2018)}]{Haji-Akbari2018Forward-fluxParameters}%
  \BibitemOpen
  \bibfield  {author} {\bibinfo {author} {\bibfnamefont {A.}~\bibnamefont
  {Haji-Akbari}},\ }\href {https://doi.org/10.1063/1.5018303} {\bibfield
  {journal} {\bibinfo  {journal} {J. Chem. Phys.}\ }\textbf {\bibinfo {volume}
  {149}},\ \bibinfo {pages} {072303} (\bibinfo {year} {2018})}\BibitemShut
  {NoStop}%
\bibitem [{\citenamefont {Shi}\ and\ \citenamefont
  {Johnson}(2001)}]{Shi2001HistogramFluids}%
  \BibitemOpen
  \bibfield  {author} {\bibinfo {author} {\bibfnamefont {W.}~\bibnamefont
  {Shi}}\ and\ \bibinfo {author} {\bibfnamefont {J.~K.}\ \bibnamefont
  {Johnson}},\ }\href {https://doi.org/10.1016/S0378-3812(01)00534-9}
  {\bibfield  {journal} {\bibinfo  {journal} {Fluid Phase Equilib.}\ }\textbf
  {\bibinfo {volume} {187-188}},\ \bibinfo {pages} {171} (\bibinfo {year}
  {2001})}\BibitemShut {NoStop}%
\bibitem [{\citenamefont {Johnson}\ \emph {et~al.}(1993)\citenamefont
  {Johnson}, \citenamefont {Zollweg},\ and\ \citenamefont
  {Gubbins}}]{Johnson1993TheRevisited}%
  \BibitemOpen
  \bibfield  {author} {\bibinfo {author} {\bibfnamefont {J.~K.}\ \bibnamefont
  {Johnson}}, \bibinfo {author} {\bibfnamefont {J.~A.}\ \bibnamefont
  {Zollweg}},\ and\ \bibinfo {author} {\bibfnamefont {K.~E.}\ \bibnamefont
  {Gubbins}},\ }\href {https://doi.org/10.1080/00268979300100411} {\bibfield
  {journal} {\bibinfo  {journal} {Mol. Phys.}\ }\textbf {\bibinfo {volume}
  {78}},\ \bibinfo {pages} {591} (\bibinfo {year} {1993})}\BibitemShut
  {NoStop}%
\bibitem [{\citenamefont {Potoff}\ and\ \citenamefont
  {Panagiotopoulos}(1998)}]{Potoff1998CriticalMixture}%
  \BibitemOpen
  \bibfield  {author} {\bibinfo {author} {\bibfnamefont {J.~J.}\ \bibnamefont
  {Potoff}}\ and\ \bibinfo {author} {\bibfnamefont {A.~Z.}\ \bibnamefont
  {Panagiotopoulos}},\ }\href {https://doi.org/10.1063/1.477787} {\bibfield
  {journal} {\bibinfo  {journal} {J. Chem. Phys.}\ }\textbf {\bibinfo {volume}
  {109}},\ \bibinfo {pages} {10914} (\bibinfo {year} {1998})}\BibitemShut
  {NoStop}%
\bibitem [{\citenamefont {Thompson}\ \emph {et~al.}(2022)\citenamefont
  {Thompson}, \citenamefont {Aktulga}, \citenamefont {Berger}, \citenamefont
  {Bolintineanu}, \citenamefont {Brown}, \citenamefont {Crozier}, \citenamefont
  {in~'t Veld}, \citenamefont {Kohlmeyer}, \citenamefont {Moore}, \citenamefont
  {Nguyen}, \citenamefont {Shan}, \citenamefont {Stevens}, \citenamefont
  {Tranchida}, \citenamefont {Trott},\ and\ \citenamefont
  {Plimpton}}]{Thompson2022LAMMPSScales}%
  \BibitemOpen
  \bibfield  {author} {\bibinfo {author} {\bibfnamefont {A.~P.}\ \bibnamefont
  {Thompson}}, \bibinfo {author} {\bibfnamefont {H.~M.}\ \bibnamefont
  {Aktulga}}, \bibinfo {author} {\bibfnamefont {R.}~\bibnamefont {Berger}},
  \bibinfo {author} {\bibfnamefont {D.~S.}\ \bibnamefont {Bolintineanu}},
  \bibinfo {author} {\bibfnamefont {W.~M.}\ \bibnamefont {Brown}}, \bibinfo
  {author} {\bibfnamefont {P.~S.}\ \bibnamefont {Crozier}}, \bibinfo {author}
  {\bibfnamefont {P.~J.}\ \bibnamefont {in~'t Veld}}, \bibinfo {author}
  {\bibfnamefont {A.}~\bibnamefont {Kohlmeyer}}, \bibinfo {author}
  {\bibfnamefont {S.~G.}\ \bibnamefont {Moore}}, \bibinfo {author}
  {\bibfnamefont {T.~D.}\ \bibnamefont {Nguyen}}, \bibinfo {author}
  {\bibfnamefont {R.}~\bibnamefont {Shan}}, \bibinfo {author} {\bibfnamefont
  {M.~J.}\ \bibnamefont {Stevens}}, \bibinfo {author} {\bibfnamefont
  {J.}~\bibnamefont {Tranchida}}, \bibinfo {author} {\bibfnamefont
  {C.}~\bibnamefont {Trott}},\ and\ \bibinfo {author} {\bibfnamefont {S.~J.}\
  \bibnamefont {Plimpton}},\ }\href {https://doi.org/10.1016/J.CPC.2021.108171}
  {\bibfield  {journal} {\bibinfo  {journal} {Comput. Phys. Commun.}\ }\textbf
  {\bibinfo {volume} {271}},\ \bibinfo {pages} {108171} (\bibinfo {year}
  {2022})}\BibitemShut {NoStop}%
\bibitem [{\citenamefont {Swope}\ \emph {et~al.}(1982)\citenamefont {Swope},
  \citenamefont {Andersen}, \citenamefont {Berens},\ and\ \citenamefont
  {Wilson}}]{SwopeJCP1982}%
  \BibitemOpen
  \bibfield  {author} {\bibinfo {author} {\bibfnamefont {W.~C.}\ \bibnamefont
  {Swope}}, \bibinfo {author} {\bibfnamefont {H.~C.}\ \bibnamefont {Andersen}},
  \bibinfo {author} {\bibfnamefont {P.~H.}\ \bibnamefont {Berens}},\ and\
  \bibinfo {author} {\bibfnamefont {K.~R.}\ \bibnamefont {Wilson}},\ }\href
  {https://dx.doi.org/10.1063/1.442716} {\bibfield  {journal} {\bibinfo
  {journal} {J. Chem. Phys.}\ }\textbf {\bibinfo {volume} {76}},\ \bibinfo
  {pages} {637} (\bibinfo {year} {1982})}\BibitemShut {NoStop}%
\bibitem [{\citenamefont {Nos\'{e}}(1984)}]{Nose1984AEnsemble}%
  \BibitemOpen
  \bibfield  {author} {\bibinfo {author} {\bibfnamefont {S.}~\bibnamefont
  {Nos\'{e}}},\ }\href {https://doi.org/10.1080/00268978400101201} {\bibfield
  {journal} {\bibinfo  {journal} {Mol. Phys.}\ }\textbf {\bibinfo {volume}
  {52}},\ \bibinfo {pages} {255} (\bibinfo {year} {1984})}\BibitemShut
  {NoStop}%
\bibitem [{\citenamefont {Hoover}(1985)}]{Hoover1985CanonicalDistributions}%
  \BibitemOpen
  \bibfield  {author} {\bibinfo {author} {\bibfnamefont {W.~G.}\ \bibnamefont
  {Hoover}},\ }\href {https://doi.org/10.1103/PhysRevA.31.1695} {\bibfield
  {journal} {\bibinfo  {journal} {Phys. Rev. A}\ }\textbf {\bibinfo {volume}
  {31}},\ \bibinfo {pages} {1695} (\bibinfo {year} {1985})}\BibitemShut
  {NoStop}%
\bibitem [{\citenamefont {Parrinello}\ and\ \citenamefont
  {Rahman}(1981)}]{Parrinello1981PolymorphicMethod}%
  \BibitemOpen
  \bibfield  {author} {\bibinfo {author} {\bibfnamefont {M.}~\bibnamefont
  {Parrinello}}\ and\ \bibinfo {author} {\bibfnamefont {A.}~\bibnamefont
  {Rahman}},\ }\href {https://doi.org/10.1063/1.328693} {\bibfield  {journal}
  {\bibinfo  {journal} {J. Appl. Phys.}\ }\textbf {\bibinfo {volume} {52}},\
  \bibinfo {pages} {7182} (\bibinfo {year} {1981})}\BibitemShut {NoStop}%
\bibitem [{\citenamefont {Frenkel}\ and\ \citenamefont
  {Ladd}(1984)}]{Frenkel1984NewSpheres}%
  \BibitemOpen
  \bibfield  {author} {\bibinfo {author} {\bibfnamefont {D.}~\bibnamefont
  {Frenkel}}\ and\ \bibinfo {author} {\bibfnamefont {A.~J.~C.}\ \bibnamefont
  {Ladd}},\ }\href {https://doi.org/10.1063/1.448024} {\bibfield  {journal}
  {\bibinfo  {journal} {J. Chem. Phys.}\ }\textbf {\bibinfo {volume} {81}},\
  \bibinfo {pages} {3188} (\bibinfo {year} {1984})}\BibitemShut {NoStop}%
\bibitem [{\citenamefont {Haji-Akbari}\ \emph {et~al.}(2011)\citenamefont
  {Haji-Akbari}, \citenamefont {Engel},\ and\ \citenamefont
  {Glotzer}}]{Haji-Akbari2011PhaseTetrahedra}%
  \BibitemOpen
  \bibfield  {author} {\bibinfo {author} {\bibfnamefont {A.}~\bibnamefont
  {Haji-Akbari}}, \bibinfo {author} {\bibfnamefont {M.}~\bibnamefont {Engel}},\
  and\ \bibinfo {author} {\bibfnamefont {S.~C.}\ \bibnamefont {Glotzer}},\
  }\href {https://doi.org/10.1063/1.3651370} {\bibfield  {journal} {\bibinfo
  {journal} {J. Chem. Phys.}\ }\textbf {\bibinfo {volume} {135}},\ \bibinfo
  {pages} {194101} (\bibinfo {year} {2011})}\BibitemShut {NoStop}%
\bibitem [{\citenamefont {Sulantay~Vargas}\ \emph {et~al.}(2026)\citenamefont
  {Sulantay~Vargas}, \citenamefont {Hussain},\ and\ \citenamefont
  {Haji-Akbari}}]{SulantayVargas2026}%
  \BibitemOpen
  \bibfield  {author} {\bibinfo {author} {\bibfnamefont {F.}~\bibnamefont
  {Sulantay~Vargas}}, \bibinfo {author} {\bibfnamefont {S.}~\bibnamefont
  {Hussain}},\ and\ \bibinfo {author} {\bibfnamefont {A.}~\bibnamefont
  {Haji-Akbari}},\ }\href {https://doi.org/10.1021/acs.cgd.5c01399} {\bibfield
  {journal} {\bibinfo  {journal} {Cryst. Growth Des.}\ }\textbf {\bibinfo
  {volume} {26}},\ \bibinfo {pages} {1943} (\bibinfo {year}
  {2026})}\BibitemShut {NoStop}%
\bibitem [{\citenamefont {Martyna}\ \emph {et~al.}(1992)\citenamefont
  {Martyna}, \citenamefont {Klein},\ and\ \citenamefont
  {Tuckerman}}]{Martyna1992Nose-HooverDynamics}%
  \BibitemOpen
  \bibfield  {author} {\bibinfo {author} {\bibfnamefont {G.~J.}\ \bibnamefont
  {Martyna}}, \bibinfo {author} {\bibfnamefont {M.~L.}\ \bibnamefont {Klein}},\
  and\ \bibinfo {author} {\bibfnamefont {M.}~\bibnamefont {Tuckerman}},\ }\href
  {https://doi.org/10.1063/1.463940} {\bibfield  {journal} {\bibinfo  {journal}
  {J. Chem. Phys.}\ }\textbf {\bibinfo {volume} {97}},\ \bibinfo {pages} {2635}
  (\bibinfo {year} {1992})}\BibitemShut {NoStop}%
\bibitem [{\citenamefont {Allen}\ \emph
  {et~al.}(2006{\natexlab{a}})\citenamefont {Allen}, \citenamefont {Frenkel},\
  and\ \citenamefont {Ten~Wolde}}]{Allen2006SimulatingSystems}%
  \BibitemOpen
  \bibfield  {author} {\bibinfo {author} {\bibfnamefont {R.~J.}\ \bibnamefont
  {Allen}}, \bibinfo {author} {\bibfnamefont {D.}~\bibnamefont {Frenkel}},\
  and\ \bibinfo {author} {\bibfnamefont {P.~R.}\ \bibnamefont {Ten~Wolde}},\
  }\href {https://doi.org/10.1063/1.2140273} {\bibfield  {journal} {\bibinfo
  {journal} {J. Chem. Phys.}\ }\textbf {\bibinfo {volume} {124}},\ \bibinfo
  {pages} {024102} (\bibinfo {year} {2006}{\natexlab{a}})}\BibitemShut
  {NoStop}%
\bibitem [{\citenamefont {Hussain}\ and\ \citenamefont
  {Haji-Akbari}(2020)}]{Hussain2020StudyingOutlook}%
  \BibitemOpen
  \bibfield  {author} {\bibinfo {author} {\bibfnamefont {S.}~\bibnamefont
  {Hussain}}\ and\ \bibinfo {author} {\bibfnamefont {A.}~\bibnamefont
  {Haji-Akbari}},\ }\href {https://doi.org/10.1063/1.5127780} {\bibfield
  {journal} {\bibinfo  {journal} {J. Chem. Phys.}\ }\textbf {\bibinfo {volume}
  {152}},\ \bibinfo {pages} {060901} (\bibinfo {year} {2020})}\BibitemShut
  {NoStop}%
\bibitem [{\citenamefont {Lechner}\ and\ \citenamefont
  {Dellago}(2008)}]{Lechner2008AccurateParameters}%
  \BibitemOpen
  \bibfield  {author} {\bibinfo {author} {\bibfnamefont {W.}~\bibnamefont
  {Lechner}}\ and\ \bibinfo {author} {\bibfnamefont {C.}~\bibnamefont
  {Dellago}},\ }\href {https://doi.org/10.1063/1.2977970} {\bibfield  {journal}
  {\bibinfo  {journal} {J. Chem. Phys.}\ }\textbf {\bibinfo {volume} {129}},\
  \bibinfo {pages} {114707} (\bibinfo {year} {2008})}\BibitemShut {NoStop}%
\bibitem [{\citenamefont {Steinhardt}\ \emph {et~al.}(1983)\citenamefont
  {Steinhardt}, \citenamefont {Nelson},\ and\ \citenamefont
  {Ronchetti}}]{Steinhardt1983Bond-orientationalGlasses}%
  \BibitemOpen
  \bibfield  {author} {\bibinfo {author} {\bibfnamefont {P.~J.}\ \bibnamefont
  {Steinhardt}}, \bibinfo {author} {\bibfnamefont {D.~R.}\ \bibnamefont
  {Nelson}},\ and\ \bibinfo {author} {\bibfnamefont {M.}~\bibnamefont
  {Ronchetti}},\ }\href {https://doi.org/10.1103/PhysRevB.28.784} {\bibfield
  {journal} {\bibinfo  {journal} {Phys. Rev. B}\ }\textbf {\bibinfo {volume}
  {28}},\ \bibinfo {pages} {784} (\bibinfo {year} {1983})}\BibitemShut
  {NoStop}%
\bibitem [{\citenamefont {Allen}\ \emph
  {et~al.}(2006{\natexlab{b}})\citenamefont {Allen}, \citenamefont {Frenkel},\
  and\ \citenamefont {Ten~Wolde}}]{AllenJCP2006}%
  \BibitemOpen
  \bibfield  {author} {\bibinfo {author} {\bibfnamefont {R.~J.}\ \bibnamefont
  {Allen}}, \bibinfo {author} {\bibfnamefont {D.}~\bibnamefont {Frenkel}},\
  and\ \bibinfo {author} {\bibfnamefont {P.~R.}\ \bibnamefont {Ten~Wolde}},\
  }\href {https://doi.org/10.1063/1.2198827} {\bibfield  {journal} {\bibinfo
  {journal} {J. Chem. Phys.}\ }\textbf {\bibinfo {volume} {124}},\ \bibinfo
  {pages} {194111} (\bibinfo {year} {2006}{\natexlab{b}})}\BibitemShut
  {NoStop}%
\bibitem [{\citenamefont {Agrawal}\ and\ \citenamefont
  {Kofke}(1995)}]{Agrawal1995ThermodynamicCoexistence}%
  \BibitemOpen
  \bibfield  {author} {\bibinfo {author} {\bibfnamefont {R.}~\bibnamefont
  {Agrawal}}\ and\ \bibinfo {author} {\bibfnamefont {D.~A.}\ \bibnamefont
  {Kofke}},\ }\href {https://doi.org/10.1080/00268979500100921} {\bibfield
  {journal} {\bibinfo  {journal} {Mol. Phys.}\ }\textbf {\bibinfo {volume}
  {85}},\ \bibinfo {pages} {43} (\bibinfo {year} {1995})}\BibitemShut {NoStop}%
\bibitem [{\citenamefont {Stillinger}\ and\ \citenamefont
  {Weber}(1985)}]{StillingerPRB1985}%
  \BibitemOpen
  \bibfield  {author} {\bibinfo {author} {\bibfnamefont {F.~H.}\ \bibnamefont
  {Stillinger}}\ and\ \bibinfo {author} {\bibfnamefont {T.~A.}\ \bibnamefont
  {Weber}},\ }\href {https://doi.org/10.1103/PhysRevB.31.5262} {\bibfield
  {journal} {\bibinfo  {journal} {Phys. Rev. B}\ }\textbf {\bibinfo {volume}
  {31}},\ \bibinfo {pages} {5262} (\bibinfo {year} {1985})}\BibitemShut
  {NoStop}%
\bibitem [{\citenamefont {Molinero}\ and\ \citenamefont
  {Moore}(2009)}]{MolineroJPCB2009}%
  \BibitemOpen
  \bibfield  {author} {\bibinfo {author} {\bibfnamefont {V.}~\bibnamefont
  {Molinero}}\ and\ \bibinfo {author} {\bibfnamefont {E.~B.}\ \bibnamefont
  {Moore}},\ }\href {https://doi.org/10.1021/jp805227c} {\bibfield  {journal}
  {\bibinfo  {journal} {J. Phys. Chem. B}\ }\textbf {\bibinfo {volume} {113}},\
  \bibinfo {pages} {4008} (\bibinfo {year} {2009})}\BibitemShut {NoStop}%
\bibitem [{\citenamefont {Zwanzig}(1954)}]{ZwanzigJCP1954}%
  \BibitemOpen
  \bibfield  {author} {\bibinfo {author} {\bibfnamefont {R.~W.}\ \bibnamefont
  {Zwanzig}},\ }\href {https://doi.org/10.1063/1.1740409} {\bibfield  {journal}
  {\bibinfo  {journal} {J. Chem. Phys.}\ }\textbf {\bibinfo {volume} {22}},\
  \bibinfo {pages} {1420} (\bibinfo {year} {1954})}\BibitemShut {NoStop}%
\bibitem [{\citenamefont {Bi}\ and\ \citenamefont {Li}(2014)}]{BiJPCB2014}%
  \BibitemOpen
  \bibfield  {author} {\bibinfo {author} {\bibfnamefont {Y.}~\bibnamefont
  {Bi}}\ and\ \bibinfo {author} {\bibfnamefont {T.}~\bibnamefont {Li}},\ }\href
  {https://doi.org/10.1021/jp503000u} {\bibfield  {journal} {\bibinfo
  {journal} {J. Phys. Chem. B}\ }\textbf {\bibinfo {volume} {118}},\ \bibinfo
  {pages} {13324} (\bibinfo {year} {2014})}\BibitemShut {NoStop}%
\bibitem [{\citenamefont {Bamer}\ \emph {et~al.}(2023)\citenamefont {Bamer},
  \citenamefont {Ebrahem}, \citenamefont {Markert},\ and\ \citenamefont
  {Stamm}}]{BamerArchComputMethodsEng2023}%
  \BibitemOpen
  \bibfield  {author} {\bibinfo {author} {\bibfnamefont {F.}~\bibnamefont
  {Bamer}}, \bibinfo {author} {\bibfnamefont {F.}~\bibnamefont {Ebrahem}},
  \bibinfo {author} {\bibfnamefont {B.}~\bibnamefont {Markert}},\ and\ \bibinfo
  {author} {\bibfnamefont {B.}~\bibnamefont {Stamm}},\ }\href
  {https://doi.org/10.1007/s11831-022-09861-1} {\bibfield  {journal} {\bibinfo
  {journal} {Arch. Comput. Methods Eng.}\ }\textbf {\bibinfo {volume} {30}},\
  \bibinfo {pages} {2105} (\bibinfo {year} {2023})}\BibitemShut {NoStop}%
\bibitem [{\citenamefont {Vehkam{\"a}ki}\ and\ \citenamefont
  {Ford}(2000)}]{VehkamkiJCP2000}%
  \BibitemOpen
  \bibfield  {author} {\bibinfo {author} {\bibfnamefont {H.}~\bibnamefont
  {Vehkam{\"a}ki}}\ and\ \bibinfo {author} {\bibfnamefont {I.~J.}\ \bibnamefont
  {Ford}},\ }\href {https://doi.org/10.1063/1.480964} {\bibfield  {journal}
  {\bibinfo  {journal} {J. Chem. Phys.}\ }\textbf {\bibinfo {volume} {112}},\
  \bibinfo {pages} {4193} (\bibinfo {year} {2000})}\BibitemShut {NoStop}%
\bibitem [{\citenamefont {Bradley}(1932)}]{BradleyPhil1932}%
  \BibitemOpen
  \bibfield  {author} {\bibinfo {author} {\bibfnamefont {R.~S.}\ \bibnamefont
  {Bradley}},\ }\href {https://doi.org/10.1080/14786449209461990} {\bibfield
  {journal} {\bibinfo  {journal} {Lond. Edinb. Dubl. Philos. Mag.}\ }\textbf
  {\bibinfo {volume} {13}},\ \bibinfo {pages} {853} (\bibinfo {year}
  {1932})}\BibitemShut {NoStop}%
\bibitem [{\citenamefont {Lag\"{u}e}\ \emph {et~al.}(2004)\citenamefont
  {Lag\"{u}e}, \citenamefont {Pastor},\ and\ \citenamefont
  {Brooks}}]{LagueJPCB2004}%
  \BibitemOpen
  \bibfield  {author} {\bibinfo {author} {\bibfnamefont {P.}~\bibnamefont
  {Lag\"{u}e}}, \bibinfo {author} {\bibfnamefont {R.~W.}\ \bibnamefont
  {Pastor}},\ and\ \bibinfo {author} {\bibfnamefont {B.~R.}\ \bibnamefont
  {Brooks}},\ }\href {https://doi.org/10.1021/jp030458y} {\bibfield  {journal}
  {\bibinfo  {journal} {J. Phys. Chem. B}\ }\textbf {\bibinfo {volume} {108}},\
  \bibinfo {pages} {363} (\bibinfo {year} {2004})}\BibitemShut {NoStop}%
\bibitem [{\citenamefont {Wennberg}\ \emph {et~al.}(2013)\citenamefont
  {Wennberg}, \citenamefont {Murtola}, \citenamefont {Hess},\ and\
  \citenamefont {Lindahl}}]{WennbergJCTC2013}%
  \BibitemOpen
  \bibfield  {author} {\bibinfo {author} {\bibfnamefont {C.~L.}\ \bibnamefont
  {Wennberg}}, \bibinfo {author} {\bibfnamefont {T.}~\bibnamefont {Murtola}},
  \bibinfo {author} {\bibfnamefont {B.}~\bibnamefont {Hess}},\ and\ \bibinfo
  {author} {\bibfnamefont {E.}~\bibnamefont {Lindahl}},\ }\href
  {https://doi.org/10.1021/ct400140n} {\bibfield  {journal} {\bibinfo
  {journal} {J. Chem. Theory Comput.}\ }\textbf {\bibinfo {volume} {9}},\
  \bibinfo {pages} {3527} (\bibinfo {year} {2013})}\BibitemShut {NoStop}%
\bibitem [{\citenamefont {Fischer}\ \emph {et~al.}(2015)\citenamefont
  {Fischer}, \citenamefont {van Maaren}, \citenamefont {Ditz}, \citenamefont
  {Yildirim},\ and\ \citenamefont {van~der Spoel}}]{FischerJCTC2015}%
  \BibitemOpen
  \bibfield  {author} {\bibinfo {author} {\bibfnamefont {N.~M.}\ \bibnamefont
  {Fischer}}, \bibinfo {author} {\bibfnamefont {P.~J.}\ \bibnamefont {van
  Maaren}}, \bibinfo {author} {\bibfnamefont {J.~C.}\ \bibnamefont {Ditz}},
  \bibinfo {author} {\bibfnamefont {A.}~\bibnamefont {Yildirim}},\ and\
  \bibinfo {author} {\bibfnamefont {D.}~\bibnamefont {van~der Spoel}},\ }\href
  {https://doi.org/10.1021/acs.jctc.5b00190} {\bibfield  {journal} {\bibinfo
  {journal} {J. Chem. Theory Comput.}\ }\textbf {\bibinfo {volume} {11}},\
  \bibinfo {pages} {2938} (\bibinfo {year} {2015})}\BibitemShut {NoStop}%
\bibitem [{\citenamefont {Atherton}\ \emph {et~al.}(2022)\citenamefont
  {Atherton}, \citenamefont {Michaelides},\ and\ \citenamefont
  {Cox}}]{Atherton2022}%
  \BibitemOpen
  \bibfield  {author} {\bibinfo {author} {\bibfnamefont {D.}~\bibnamefont
  {Atherton}}, \bibinfo {author} {\bibfnamefont {A.}~\bibnamefont
  {Michaelides}},\ and\ \bibinfo {author} {\bibfnamefont {S.~J.}\ \bibnamefont
  {Cox}},\ }\href {https://doi.org/10.1063/5.0085750} {\bibfield  {journal}
  {\bibinfo  {journal} {J. Chem. Phys.}\ }\textbf {\bibinfo {volume} {156}},\
  \bibinfo {pages} {164501} (\bibinfo {year} {2022})}\BibitemShut {NoStop}%
\bibitem [{\citenamefont {Horn}\ \emph {et~al.}(2004)\citenamefont {Horn},
  \citenamefont {Swope}, \citenamefont {Pitera}, \citenamefont {Madura},
  \citenamefont {Dick}, \citenamefont {Hura},\ and\ \citenamefont
  {Head-Gordon}}]{HornJCP2004}%
  \BibitemOpen
  \bibfield  {author} {\bibinfo {author} {\bibfnamefont {H.~W.}\ \bibnamefont
  {Horn}}, \bibinfo {author} {\bibfnamefont {W.~C.}\ \bibnamefont {Swope}},
  \bibinfo {author} {\bibfnamefont {J.~W.}\ \bibnamefont {Pitera}}, \bibinfo
  {author} {\bibfnamefont {J.~D.}\ \bibnamefont {Madura}}, \bibinfo {author}
  {\bibfnamefont {T.~J.}\ \bibnamefont {Dick}}, \bibinfo {author}
  {\bibfnamefont {G.~L.}\ \bibnamefont {Hura}},\ and\ \bibinfo {author}
  {\bibfnamefont {T.}~\bibnamefont {Head-Gordon}},\ }\href
  {https://dx.doi.org/10.1063/1.1683075} {\bibfield  {journal} {\bibinfo
  {journal} {J. Chem. Phys.}\ }\textbf {\bibinfo {volume} {120}},\ \bibinfo
  {pages} {9665} (\bibinfo {year} {2004})}\BibitemShut {NoStop}%
\bibitem [{\citenamefont {Abascal}\ and\ \citenamefont
  {Vega}(2005)}]{AbascalJCP2005}%
  \BibitemOpen
  \bibfield  {author} {\bibinfo {author} {\bibfnamefont {J.~L.~F.}\
  \bibnamefont {Abascal}}\ and\ \bibinfo {author} {\bibfnamefont
  {C.}~\bibnamefont {Vega}},\ }\href {https://doi.org/10.1063/1.2121687}
  {\bibfield  {journal} {\bibinfo  {journal} {J. Chem. Phys.}\ }\textbf
  {\bibinfo {volume} {123}},\ \bibinfo {pages} {234505} (\bibinfo {year}
  {2005})}\BibitemShut {NoStop}%
\bibitem [{\citenamefont {Abascal}\ \emph {et~al.}(2005)\citenamefont
  {Abascal}, \citenamefont {Sanz}, \citenamefont {Garc{\'\i}a~Fern{\'a}ndez},\
  and\ \citenamefont {Vega}}]{AbascalJCP2005p}%
  \BibitemOpen
  \bibfield  {author} {\bibinfo {author} {\bibfnamefont {J.}~\bibnamefont
  {Abascal}}, \bibinfo {author} {\bibfnamefont {E.}~\bibnamefont {Sanz}},
  \bibinfo {author} {\bibfnamefont {R.}~\bibnamefont
  {Garc{\'\i}a~Fern{\'a}ndez}},\ and\ \bibinfo {author} {\bibfnamefont
  {C.}~\bibnamefont {Vega}},\ }\href {https://doi.org/10.1063/1.1931662}
  {\bibfield  {journal} {\bibinfo  {journal} {J. Chem. Phys.}\ }\textbf
  {\bibinfo {volume} {122}},\ \bibinfo {pages} {234511} (\bibinfo {year}
  {2005})}\BibitemShut {NoStop}%
\bibitem [{\citenamefont {Jorgensen}\ \emph {et~al.}(1996)\citenamefont
  {Jorgensen}, \citenamefont {Maxwell},\ and\ \citenamefont
  {Tirado-Rives}}]{JorgensenJACS1996}%
  \BibitemOpen
  \bibfield  {author} {\bibinfo {author} {\bibfnamefont {W.~L.}\ \bibnamefont
  {Jorgensen}}, \bibinfo {author} {\bibfnamefont {D.~S.}\ \bibnamefont
  {Maxwell}},\ and\ \bibinfo {author} {\bibfnamefont {J.}~\bibnamefont
  {Tirado-Rives}},\ }\href {https://doi.org/10.1021/ja9621760} {\bibfield
  {journal} {\bibinfo  {journal} {J. Am. Chem. Soc.}\ }\textbf {\bibinfo
  {volume} {118}},\ \bibinfo {pages} {11225} (\bibinfo {year}
  {1996})}\BibitemShut {NoStop}%
\bibitem [{\citenamefont {Martin}\ and\ \citenamefont
  {Siepmann}(1998)}]{Martin1998}%
  \BibitemOpen
  \bibfield  {author} {\bibinfo {author} {\bibfnamefont {M.~G.}\ \bibnamefont
  {Martin}}\ and\ \bibinfo {author} {\bibfnamefont {J.~I.}\ \bibnamefont
  {Siepmann}},\ }\href {https://doi.org/10.1021/jp972543+} {\bibfield
  {journal} {\bibinfo  {journal} {J. Phys. Chem. B}\ }\textbf {\bibinfo
  {volume} {102}},\ \bibinfo {pages} {2569} (\bibinfo {year}
  {1998})}\BibitemShut {NoStop}%
\bibitem [{\citenamefont {Sosso}\ \emph
  {et~al.}(2016{\natexlab{b}})\citenamefont {Sosso}, \citenamefont {Tribello},
  \citenamefont {Zen}, \citenamefont {Pedevilla},\ and\ \citenamefont
  {Michaelides}}]{Sosso2016}%
  \BibitemOpen
  \bibfield  {author} {\bibinfo {author} {\bibfnamefont {G.~C.}\ \bibnamefont
  {Sosso}}, \bibinfo {author} {\bibfnamefont {G.~A.}\ \bibnamefont {Tribello}},
  \bibinfo {author} {\bibfnamefont {A.}~\bibnamefont {Zen}}, \bibinfo {author}
  {\bibfnamefont {P.}~\bibnamefont {Pedevilla}},\ and\ \bibinfo {author}
  {\bibfnamefont {A.}~\bibnamefont {Michaelides}},\ }\href
  {https://doi.org/10.1063/1.4968796} {\bibfield  {journal} {\bibinfo
  {journal} {J. Chem. Phys.}\ }\textbf {\bibinfo {volume} {145}},\ \bibinfo
  {pages} {211927} (\bibinfo {year} {2016}{\natexlab{b}})}\BibitemShut
  {NoStop}%
\bibitem [{\citenamefont {Fitzner}\ \emph {et~al.}(2017)\citenamefont
  {Fitzner}, \citenamefont {Sosso}, \citenamefont {Pietrucci}, \citenamefont
  {Pipolo},\ and\ \citenamefont {Michaelides}}]{Fitzner2017}%
  \BibitemOpen
  \bibfield  {author} {\bibinfo {author} {\bibfnamefont {M.}~\bibnamefont
  {Fitzner}}, \bibinfo {author} {\bibfnamefont {G.~C.}\ \bibnamefont {Sosso}},
  \bibinfo {author} {\bibfnamefont {F.}~\bibnamefont {Pietrucci}}, \bibinfo
  {author} {\bibfnamefont {S.}~\bibnamefont {Pipolo}},\ and\ \bibinfo {author}
  {\bibfnamefont {A.}~\bibnamefont {Michaelides}},\ }\href
  {https://doi.org/10.1038/s41467-017-02300-x} {\bibfield  {journal} {\bibinfo
  {journal} {Nat. Commun.}\ }\textbf {\bibinfo {volume} {8}},\ \bibinfo {pages}
  {2257} (\bibinfo {year} {2017})}\BibitemShut {NoStop}%
\bibitem [{\citenamefont {de~Leeuw}\ \emph
  {et~al.}(1980{\natexlab{a}})\citenamefont {de~Leeuw}, \citenamefont
  {Perram},\ and\ \citenamefont {Smith}}]{deLeeuwPRSLA1980}%
  \BibitemOpen
  \bibfield  {author} {\bibinfo {author} {\bibfnamefont {S.~W.}\ \bibnamefont
  {de~Leeuw}}, \bibinfo {author} {\bibfnamefont {J.~W.}\ \bibnamefont
  {Perram}},\ and\ \bibinfo {author} {\bibfnamefont {E.~R.}\ \bibnamefont
  {Smith}},\ }\href {https://dx.doi.org/10.1098/rspa.1980.0135} {\bibfield
  {journal} {\bibinfo  {journal} {Proc. R. Soc. Lond. A Math. Phys. Sci.}\
  }\textbf {\bibinfo {volume} {373}},\ \bibinfo {pages} {27} (\bibinfo {year}
  {1980}{\natexlab{a}})}\BibitemShut {NoStop}%
\bibitem [{\citenamefont {de~Leeuw}\ \emph
  {et~al.}(1980{\natexlab{b}})\citenamefont {de~Leeuw}, \citenamefont
  {Perram},\ and\ \citenamefont {Smith}}]{LdeeeuwPRSLA1980p}%
  \BibitemOpen
  \bibfield  {author} {\bibinfo {author} {\bibfnamefont {S.~W.}\ \bibnamefont
  {de~Leeuw}}, \bibinfo {author} {\bibfnamefont {J.~W.}\ \bibnamefont
  {Perram}},\ and\ \bibinfo {author} {\bibfnamefont {E.~R.}\ \bibnamefont
  {Smith}},\ }\href {https://dx.doi.org/10.1098/rspa.1980.0136} {\bibfield
  {journal} {\bibinfo  {journal} {Proc. R. Soc. Lond. A Math. Phys. Sci.}\
  }\textbf {\bibinfo {volume} {373}},\ \bibinfo {pages} {57} (\bibinfo {year}
  {1980}{\natexlab{b}})}\BibitemShut {NoStop}%
\bibitem [{\citenamefont {in~'t Veld}\ \emph {et~al.}(2007)\citenamefont {in~'t
  Veld}, \citenamefont {Ismail},\ and\ \citenamefont {Grest}}]{IntVeldJCP2007}%
  \BibitemOpen
  \bibfield  {author} {\bibinfo {author} {\bibfnamefont {P.~J.}\ \bibnamefont
  {in~'t Veld}}, \bibinfo {author} {\bibfnamefont {A.~E.}\ \bibnamefont
  {Ismail}},\ and\ \bibinfo {author} {\bibfnamefont {G.~S.}\ \bibnamefont
  {Grest}},\ }\href {https://doi.org/10.1063/1.2770730} {\bibfield  {journal}
  {\bibinfo  {journal} {J. Chem. Phys.}\ }\textbf {\bibinfo {volume} {127}},\
  \bibinfo {pages} {144711} (\bibinfo {year} {2007})}\BibitemShut {NoStop}%
\bibitem [{\citenamefont {MacKerell~Jr}\ \emph {et~al.}(1998)\citenamefont
  {MacKerell~Jr}, \citenamefont {Bashford}, \citenamefont {Bellott},
  \citenamefont {Dunbrack~Jr}, \citenamefont {Evanseck}, \citenamefont {Field},
  \citenamefont {Fischer}, \citenamefont {Gao}, \citenamefont {Guo},
  \citenamefont {Ha} \emph {et~al.}}]{MacKerellJPCB1998}%
  \BibitemOpen
  \bibfield  {author} {\bibinfo {author} {\bibfnamefont {A.~D.}\ \bibnamefont
  {MacKerell~Jr}}, \bibinfo {author} {\bibfnamefont {D.}~\bibnamefont
  {Bashford}}, \bibinfo {author} {\bibfnamefont {M.}~\bibnamefont {Bellott}},
  \bibinfo {author} {\bibfnamefont {R.~L.}\ \bibnamefont {Dunbrack~Jr}},
  \bibinfo {author} {\bibfnamefont {J.~D.}\ \bibnamefont {Evanseck}}, \bibinfo
  {author} {\bibfnamefont {M.~J.}\ \bibnamefont {Field}}, \bibinfo {author}
  {\bibfnamefont {S.}~\bibnamefont {Fischer}}, \bibinfo {author} {\bibfnamefont
  {J.}~\bibnamefont {Gao}}, \bibinfo {author} {\bibfnamefont {H.}~\bibnamefont
  {Guo}}, \bibinfo {author} {\bibfnamefont {S.}~\bibnamefont {Ha}}, \emph
  {et~al.},\ }\href {https://dx.doi.org/10.1021/jp973084f} {\bibfield
  {journal} {\bibinfo  {journal} {J. Phys. Chem. B}\ }\textbf {\bibinfo
  {volume} {102}},\ \bibinfo {pages} {3586} (\bibinfo {year}
  {1998})}\BibitemShut {NoStop}%
\bibitem [{\citenamefont {Kunz}\ \emph {et~al.}(2012)\citenamefont {Kunz},
  \citenamefont {Allison}, \citenamefont {Geerke}, \citenamefont {Horta},
  \citenamefont {H{\"u}nenberger}, \citenamefont {Riniker}, \citenamefont
  {Schmid},\ and\ \citenamefont {van Gunsteren}}]{KunzJComputChem2012}%
  \BibitemOpen
  \bibfield  {author} {\bibinfo {author} {\bibfnamefont {A.-P.~E.}\
  \bibnamefont {Kunz}}, \bibinfo {author} {\bibfnamefont {J.~R.}\ \bibnamefont
  {Allison}}, \bibinfo {author} {\bibfnamefont {D.~P.}\ \bibnamefont {Geerke}},
  \bibinfo {author} {\bibfnamefont {B.~A.}\ \bibnamefont {Horta}}, \bibinfo
  {author} {\bibfnamefont {P.~H.}\ \bibnamefont {H{\"u}nenberger}}, \bibinfo
  {author} {\bibfnamefont {S.}~\bibnamefont {Riniker}}, \bibinfo {author}
  {\bibfnamefont {N.}~\bibnamefont {Schmid}},\ and\ \bibinfo {author}
  {\bibfnamefont {W.~F.}\ \bibnamefont {van Gunsteren}},\ }\href
  {https://doi.org/10.1002/jcc.21954} {\bibfield  {journal} {\bibinfo
  {journal} {J. Comput. Chem.}\ }\textbf {\bibinfo {volume} {33}},\ \bibinfo
  {pages} {340} (\bibinfo {year} {2012})}\BibitemShut {NoStop}%
\bibitem [{\citenamefont {De~Leeuw}\ and\ \citenamefont
  {Perram}(1981)}]{DeLeeuwPhysicaA1981}%
  \BibitemOpen
  \bibfield  {author} {\bibinfo {author} {\bibfnamefont {S.~W.}\ \bibnamefont
  {De~Leeuw}}\ and\ \bibinfo {author} {\bibfnamefont {J.~W.}\ \bibnamefont
  {Perram}},\ }\href {https://doi.org/10.1016/0378-4371(81)90031-5} {\bibfield
  {journal} {\bibinfo  {journal} {Physica A: Stat. Mech. Appl.}\ }\textbf
  {\bibinfo {volume} {107}},\ \bibinfo {pages} {179} (\bibinfo {year}
  {1981})}\BibitemShut {NoStop}%
\bibitem [{\citenamefont {Neumann}(1985)}]{NeumannJCP1985}%
  \BibitemOpen
  \bibfield  {author} {\bibinfo {author} {\bibfnamefont {M.}~\bibnamefont
  {Neumann}},\ }\href {https://doi.org/10.1063/1.448553} {\bibfield  {journal}
  {\bibinfo  {journal} {J. Chem. Phys.}\ }\textbf {\bibinfo {volume} {82}},\
  \bibinfo {pages} {5663} (\bibinfo {year} {1985})}\BibitemShut {NoStop}%
\bibitem [{\citenamefont {Essmann}\ \emph {et~al.}(1995)\citenamefont
  {Essmann}, \citenamefont {Perera}, \citenamefont {Berkowitz}, \citenamefont
  {Darden}, \citenamefont {Lee},\ and\ \citenamefont
  {Pedersen}}]{EssmannJCP1995}%
  \BibitemOpen
  \bibfield  {author} {\bibinfo {author} {\bibfnamefont {U.}~\bibnamefont
  {Essmann}}, \bibinfo {author} {\bibfnamefont {L.}~\bibnamefont {Perera}},
  \bibinfo {author} {\bibfnamefont {M.~L.}\ \bibnamefont {Berkowitz}}, \bibinfo
  {author} {\bibfnamefont {T.}~\bibnamefont {Darden}}, \bibinfo {author}
  {\bibfnamefont {H.}~\bibnamefont {Lee}},\ and\ \bibinfo {author}
  {\bibfnamefont {L.~G.}\ \bibnamefont {Pedersen}},\ }\href
  {https://doi.org/10.1063/1.470117} {\bibfield  {journal} {\bibinfo  {journal}
  {J. Chem. Phys.}\ }\textbf {\bibinfo {volume} {103}},\ \bibinfo {pages}
  {8577} (\bibinfo {year} {1995})}\BibitemShut {NoStop}%
\bibitem [{\citenamefont {Yeh}\ and\ \citenamefont
  {Berkowitz}(1999)}]{YehJCP1999}%
  \BibitemOpen
  \bibfield  {author} {\bibinfo {author} {\bibfnamefont {I.-C.}\ \bibnamefont
  {Yeh}}\ and\ \bibinfo {author} {\bibfnamefont {M.~L.}\ \bibnamefont
  {Berkowitz}},\ }\href {https://doi.org/10.1063/1.479595} {\bibfield
  {journal} {\bibinfo  {journal} {J. Chem. Phys.}\ }\textbf {\bibinfo {volume}
  {111}},\ \bibinfo {pages} {3155} (\bibinfo {year} {1999})}\BibitemShut
  {NoStop}%
\bibitem [{\citenamefont {Liu}\ \emph {et~al.}(2012)\citenamefont {Liu},
  \citenamefont {Palmer}, \citenamefont {Panagiotopoulos},\ and\ \citenamefont
  {Debenedetti}}]{LiuJCP2012}%
  \BibitemOpen
  \bibfield  {author} {\bibinfo {author} {\bibfnamefont {Y.}~\bibnamefont
  {Liu}}, \bibinfo {author} {\bibfnamefont {J.~C.}\ \bibnamefont {Palmer}},
  \bibinfo {author} {\bibfnamefont {A.~Z.}\ \bibnamefont {Panagiotopoulos}},\
  and\ \bibinfo {author} {\bibfnamefont {P.~G.}\ \bibnamefont {Debenedetti}},\
  }\href {https://doi.org/10.1063/1.4769126} {\bibfield  {journal} {\bibinfo
  {journal} {J. Chem. Phys.}\ }\textbf {\bibinfo {volume} {137}},\ \bibinfo
  {pages} {214505} (\bibinfo {year} {2012})}\BibitemShut {NoStop}%
\bibitem [{\citenamefont {Stillinger}\ and\ \citenamefont
  {Rahman}(1974)}]{StillingerjCP1974}%
  \BibitemOpen
  \bibfield  {author} {\bibinfo {author} {\bibfnamefont {F.~H.}\ \bibnamefont
  {Stillinger}}\ and\ \bibinfo {author} {\bibfnamefont {A.}~\bibnamefont
  {Rahman}},\ }\href {https://doi.org/10.1063/1.1681229} {\bibfield  {journal}
  {\bibinfo  {journal} {J. Chem. Phys.}\ }\textbf {\bibinfo {volume} {60}},\
  \bibinfo {pages} {1545} (\bibinfo {year} {1974})}\BibitemShut {NoStop}%
\bibitem [{\citenamefont {Falahati}\ and\ \citenamefont
  {Haji-Akbari}(2019)}]{Falahati2019ThermodynamicallyBiology}%
  \BibitemOpen
  \bibfield  {author} {\bibinfo {author} {\bibfnamefont {H.}~\bibnamefont
  {Falahati}}\ and\ \bibinfo {author} {\bibfnamefont {A.}~\bibnamefont
  {Haji-Akbari}},\ }\href {https://doi.org/10.1039/C8SM02285B} {\bibfield
  {journal} {\bibinfo  {journal} {Soft Matter}\ }\textbf {\bibinfo {volume}
  {15}},\ \bibinfo {pages} {1135} (\bibinfo {year} {2019})}\BibitemShut
  {NoStop}%
\bibitem [{\citenamefont {Xu}\ \emph {et~al.}(2014)\citenamefont {Xu},
  \citenamefont {Ting}, \citenamefont {Kusaka},\ and\ \citenamefont
  {Wang}}]{XuAnnuRevPhysChem2014}%
  \BibitemOpen
  \bibfield  {author} {\bibinfo {author} {\bibfnamefont {X.}~\bibnamefont
  {Xu}}, \bibinfo {author} {\bibfnamefont {C.~L.}\ \bibnamefont {Ting}},
  \bibinfo {author} {\bibfnamefont {I.}~\bibnamefont {Kusaka}},\ and\ \bibinfo
  {author} {\bibfnamefont {Z.-G.}\ \bibnamefont {Wang}},\ }\href
  {https://doi.org/10.1146/annurev-physchem-032511-143750} {\bibfield
  {journal} {\bibinfo  {journal} {Annu. Rev. Phys. Chem.}\ }\textbf {\bibinfo
  {volume} {65}},\ \bibinfo {pages} {449} (\bibinfo {year} {2014})}\BibitemShut
  {NoStop}%
\bibitem [{\citenamefont {Towns}\ \emph {et~al.}(2014)\citenamefont {Towns},
  \citenamefont {Cockerill}, \citenamefont {Dahan}, \citenamefont {Foster},
  \citenamefont {Gaither}, \citenamefont {Grimshaw}, \citenamefont {Hazlewood},
  \citenamefont {Lathrop}, \citenamefont {Lifka}, \citenamefont {Peterson}
  \emph {et~al.}}]{TownsComp2014}%
  \BibitemOpen
  \bibfield  {author} {\bibinfo {author} {\bibfnamefont {J.}~\bibnamefont
  {Towns}}, \bibinfo {author} {\bibfnamefont {T.}~\bibnamefont {Cockerill}},
  \bibinfo {author} {\bibfnamefont {M.}~\bibnamefont {Dahan}}, \bibinfo
  {author} {\bibfnamefont {I.}~\bibnamefont {Foster}}, \bibinfo {author}
  {\bibfnamefont {K.}~\bibnamefont {Gaither}}, \bibinfo {author} {\bibfnamefont
  {A.}~\bibnamefont {Grimshaw}}, \bibinfo {author} {\bibfnamefont
  {V.}~\bibnamefont {Hazlewood}}, \bibinfo {author} {\bibfnamefont
  {S.}~\bibnamefont {Lathrop}}, \bibinfo {author} {\bibfnamefont
  {D.}~\bibnamefont {Lifka}}, \bibinfo {author} {\bibfnamefont {G.~D.}\
  \bibnamefont {Peterson}}, \emph {et~al.},\ }\href
  {https://doi.org/10.1109/MCSE.2014.80} {\bibfield  {journal} {\bibinfo
  {journal} {Comput. Sci. Eng.}\ }\textbf {\bibinfo {volume} {16}},\ \bibinfo
  {pages} {62} (\bibinfo {year} {2014})}\BibitemShut {NoStop}%
\bibitem [{\citenamefont {Boerner}\ \emph {et~al.}(2023)\citenamefont
  {Boerner}, \citenamefont {Deems}, \citenamefont {Furlani}, \citenamefont
  {Knuth},\ and\ \citenamefont {Towns}}]{BoernerACM2023}%
  \BibitemOpen
  \bibfield  {author} {\bibinfo {author} {\bibfnamefont {T.~J.}\ \bibnamefont
  {Boerner}}, \bibinfo {author} {\bibfnamefont {S.}~\bibnamefont {Deems}},
  \bibinfo {author} {\bibfnamefont {T.~R.}\ \bibnamefont {Furlani}}, \bibinfo
  {author} {\bibfnamefont {S.~L.}\ \bibnamefont {Knuth}},\ and\ \bibinfo
  {author} {\bibfnamefont {J.}~\bibnamefont {Towns}},\ }in\ \href
  {https://doi.org/10.1145/3569951.3597559} {\emph {\bibinfo {booktitle}
  {Practice and experience in advanced research computing 2023: Computing for
  the common good}}}\ (\bibinfo {year} {2023})\ pp.\ \bibinfo {pages}
  {173--176}\BibitemShut {NoStop}%
\end{thebibliography}%

\end{document}